\documentclass[a4paper]{spie}  %>>> use for US letter paper
\usepackage{amsmath,amsfonts,amssymb}
\usepackage{epigraph}
\usepackage{longtable}

\usepackage{lineno}
\usepackage{graphicx}
\usepackage[colorlinks=true, allcolors=blue]{hyperref}
\usepackage{eso-pic}
\usepackage{array} %Table column stuff
\newcolumntype{C}[1]{>{\centering\let\newline\\\arraybackslash\hspace{0pt}}p{#1}}
\usepackage{orcidlink} %a nice package for linking ORCIDs to authors
\usepackage[authoryear]{natbib}
\bibpunct{(}{)}{;}{a}{}{,}

\usepackage{pdflscape}
\usepackage{multirow}
\usepackage{caption}
\newcolumntype{L}[1]{>{\raggedright\arraybackslash}p{#1}}
\newcolumntype{C}[1]{>{\centering\arraybackslash}p{#1}}

\definecolor{headerblue}{RGB}{198, 224, 180}
\definecolor{lightgray}{RGB}{242, 242, 242}
\begin{document} 

\begingroup
\thispagestyle{empty}
\AddToShipoutPicture*{\put(0,0){\includegraphics[width=\paperwidth]{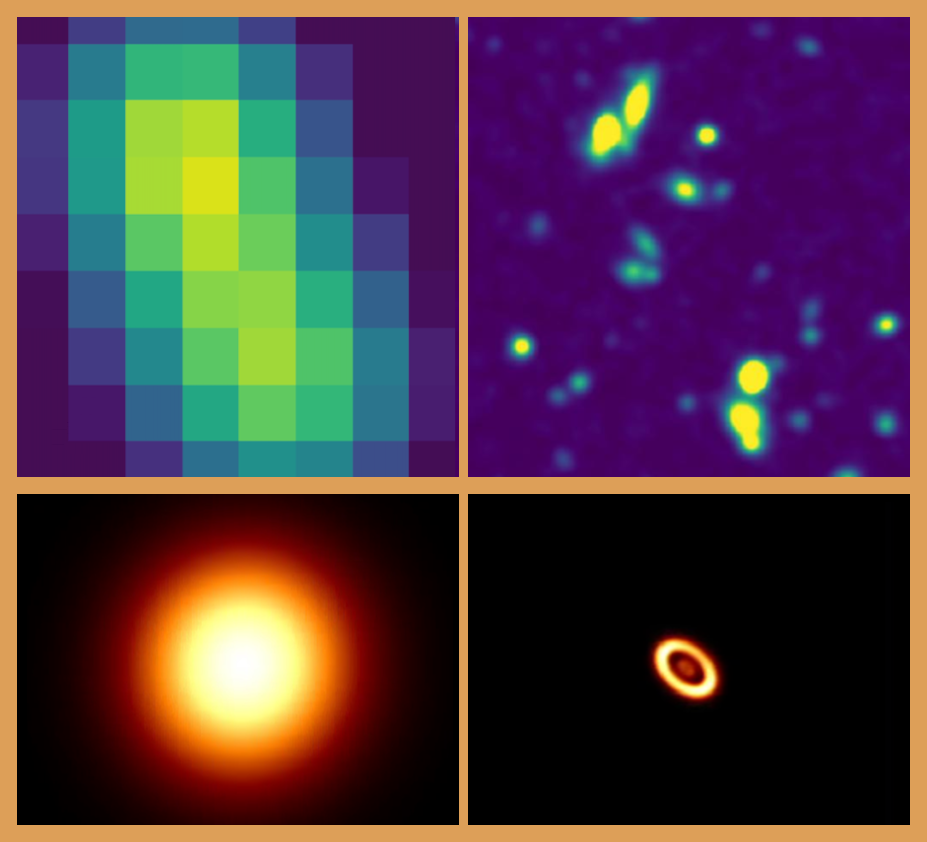}}} % Image background.
\centering
\vspace*{1cm}
\par\normalfont\fontsize{20}{20}\sffamily\selectfont
\textbf{Ushering in sub-arcsecond angular resolution observatories in the far-infrared: The case for a double-Fourier interferometer} %Something better needed?\\
{\LARGE }\par % Book title
\vspace*{1cm}
{\large{A white paper in preparation for and to accompany long-range planning. E.g., UKSA Space Frontiers 2035 (UK), NASA ASTRA 2026 (USA), and CASCA LRP 2020 midterm review (Canada).} }\par 
\vspace*{1cm}
{\large{Published October 2026}}\par
\endgroup

\newpage
{\noindent \small \textit{Cover image: Simulated observations of an extragalactic field (top) and a protoplantary disk (bottom) as measured with the level of spatial detail available to current and planned far-infrared observatories ($\sim$3.5\,m-class, left). This is compared to the capabilities of a far-infrared interferometer ($\sim$30\,m-baseline, right). We also note that a double-Fourier interferometer simultaneously retrieves spectral data for each spatial pixel shown. Images courtesy of the SPIRIT/SPICE instrument and simulation team. See Fig.~\ref{fig:scienceSummary} for a more detailed description.}}

\vspace{1em}

\renewcommand{\arraystretch}{1.5}
\noindent \textbf{Principal Author:}\\
\vspace{3em}
\begin{tabular}{l p{13.0cm}}
     Name: & C. S. Benson~\orcidlink{0000-0001-5828-5246}  \\
     Institution: & Cardiff Hub for Astrophysics Research \&\ Technology, Department of Physics \&\ Astronomy, Cardiff University, Cardiff, UK \\
     Address: & Queen’s Buildings, 14 The Parade, Cardiff CF24 3AA. UK.\\
     Email: & BensonC@Cardiff.ac.uk\\
\end{tabular}
\addtocounter{table}{-1}

\vspace{-2.5em}
\noindent \textbf{Co-authors:}
\vspace{-1em}
\begin{longtable}{l p{12.5cm}}
    M. Bonato~\orcidlink{0000-0001-9139-2342} & Instituto di Radioastronomia, Bologna, Italy; Italian ALMA Regional Centre, Bologna, Italy\\
    T. A. Davis~\orcidlink{0000-0003-4932-9379} & Cardiff Hub for Astrophysics Research \&\ Technology, Department of Physics \&\ Astronomy, Cardiff University, Cardiff, UK\\
    T. Essinger-Hileman~\orcidlink{0000-0002-4782-3851} & Observational Cosmology Laboratory, Goddard Space Flight Center, Greenbelt, MD, USA\\
    D. Farrah~\orcidlink{0000-0003-1748-2010} & Department of Physics and Astronomy, University of Hawai`i at M\=anoa, Honolulu, HI, USA; Institute for Astronomy, University of Hawai‘i, Honolulu, HI, USA \\
    W. Grainger~\orcidlink{0009-0002-9461-6107} & RAL Space, STFC Rutherford Appleton Laboratory, Didcot, Oxon, UK \\
    D. Leisawitz~\orcidlink{0000-0003-1000-2547} & Visiting Research Scientist, Department of Astronomy, University of Maryland, College Park, MD, USA\\
    H. Matsuo~\orcidlink{0000-0003-3278-2484} & National Astronomical Observatory of Japan, Mitaka, Tokyo, Japan \\
    B. Matthews & NRC Herzberg Astronomy and Astrophysics Research Centre, Victoria, BC, Canada \\
    P. Pokorny~\orcidlink{0000-0002-5667-9337} & Department of Physics, The Catholic University of America, Washington, DC, USA \\
    S. E. Ragan~\orcidlink{0000-0003-4164-5588} & Cardiff Hub for Astrophysics Research \&\ Technology, Department of Physics \&\ Astronomy, Cardiff University, Cardiff, UK\\
    B. V. Ricketti~\orcidlink{0000-0001-9701-5660} & Disruptive Space Technology Centre (DSTC), RAL Space, STFC Rutherford Appleton Laboratory, Didcot, Oxon, UK\\
    G. Savini & Department of Physics and Astronomy, University College London, Gower Street, London, UK\\
    J. P. Scott~\orcidlink{0009-0000-3882-9242} & Department of Astronomy and Steward Observatory, The University of Arizona, Tucson, AZ, USA \\
    L. D. Spencer~\orcidlink{0000-0002-9941-2077} & Institute for Space Imaging Science, Department of Physics \& Astronomy, University of Lethbridge, Lethbridge, AB, Canada\\
    C. M. Urry~\orcidlink{0000-0002-0745-9792} & Physics Department, Yale University, New Haven, CT, USA\\
    D. Wilner~\orcidlink{0000-0003-1526-7587} & Center for Astrophysics \textbar\, Harvard \& Smithsonian, Cambridge, MA, USA\\
    
\end{longtable}

\addtocounter{table}{-1}
\noindent\textbf{Additional Signatories:}
\vspace{-1em}
\begin{longtable}{l p{12.5cm}}
     E. Alei~\orcidlink{0000-0002-0006-1175} & Center for Space Sciences and Technology, University of Maryland, Baltimore County, MD, USA; NASA Goddard Space Flight Center, Greenbelt, MD, USA. \\
     M. Booth~\orcidlink{0000-0001-8568-6336} & UK Astronomy Technology Centre, Royal Observatory Edinburgh, Blackford Hill, Edinburgh, UK\\
     C. Bracken~\orcidlink{0000-0002-0773-7721} & Department of Physics, National University of Ireland, Maynooth, Ireland \\
     W. J. de Wit~\orcidlink{0000-0002-0369-5421} & European Southern Observatory, Santiago, Chile \\
     A. Duarte-Cabral~\orcidlink{0000-0002-5259-4774} &  Cardiff Hub for Astrophysics Research \&\ Technology, Department of Physics \&\ Astronomy, Cardiff University, Cardiff, UK \\
     S. Eales & Cardiff Hub for Astrophysics Research \&\ Technology, Department of Physics \&\ Astronomy, Cardiff University, Cardiff, UK \\
     S. Ertel~\orcidlink{0000-0002-2314-7289} & Department of Astronomy and Steward Observatory, The University of Arizona, Tucson, AZ, USA; Large Binocular Telescope Observatory, The University of Arizona, Tucson, AZ, USA\\
     J. Greaves & Cardiff Hub for Astrophysics Research \&\ Technology, Department of Physics \&\ Astronomy, Cardiff University, Cardiff, UK \\
     P. Hargrave~\orcidlink{0000-0002-3109-6629} & Cardiff Hub for Astrophysics Research \&\ Technology, Department of Physics \&\ Astronomy, Cardiff University, Cardiff, UK\\
     M. Kishimoto~\orcidlink{0000-0002-2216-3252} & Department of Astrophysics and Atmospheric Sciences, Kyoto Sangyo University, Japan\\
     J. B. Lovell~\orcidlink{0000-0002-4248-5443} & Center for Astrophysics \textbar\, Harvard \& Smithsonian, Cambridge, MA, USA\\
     Y. Lu~\orcidlink{0000-0002-5069-3074} & Department of Physics and Astronomy, University of Exeter, Exeter, UK \\
     J. Monnier & Department of Astronomy, University of Michigan, Ann Arbor, MI, USA\\
     L. Mundy & Department of Astronomy, University of Maryland, College Park, MD, USA \\
     J. R. Najita & National Science Foundation NOIRLab, Tucson, AZ, USA\\
     D. A. Naylor & Professor Emeritus, Department of Physics and Astronomy, University of Lethbridge, AB, Canada\\
     C. O'Sullivan~\orcidlink{0000-0001-5053-4754} & Department of Physics, National University of Ireland, Maynooth, Ireland \\
     C. Pearson~\orcidlink{0000-0001-6139-649X} & RAL Space, STFC Rutherford Appleton Laboratory, Didcot, Oxon, UK \\
     T. D. Pearce~\orcidlink{0000-0001-5653-5635} & School of Physics, Trinity College Dublin, The University of Dublin, Dublin, Ireland \\
     E. Pedretti & RAL Space, STFC Rutherford Appleton Laboratory, Didcot, Oxon, UK\\
     D. Rocha~\orcidlink{0000-0002-8225-7699} & Laborat\'{o}rio Nacional de Astrof\'{i}sica, Itajub\'{a}, MG, Brazil\\
     M. Smith & Cardiff Hub for Astrophysics Research \&\ Technology, Department of Physics \&\ Astronomy, Cardiff University, Cardiff, UK\\
     R. R. Thomson~\orcidlink{0000-0003-4978-1488} & Institute of Photonics and Quantum Sciences, Edinburgh, UK\\
     G. van Belle~\orcidlink{0000-0002-8552-158X} & Las Cumbres Observatory Global Telescope Network: Goleta, California, USA\\
     I. Veenendaal~\orcidlink{0000-0001-9488-6700} &  Cardiff Hub for Astrophysics Research \&\ Technology, Department of Physics \&\ Astronomy, Cardiff University, Cardiff, UK\\
     M. C. Wyatt~\orcidlink{0000-0001-9064-5598} & Institute of Astronomy, University of Cambridge, Cambridge, UK\\
     H. Zinnecker~\orcidlink{0000-0003-0504-3539} & Universidad Auton\'{o}ma de Chile, Santiago, Chile; NASA Ames Research Center, Moffett Field, California, USA (retired)
     
\end{longtable}

%Waiting to hear back from: Lee Mundy, Gerard van Belle, 

\newpage

\vspace{1em}

\begin{abstract}
A far-infrared (far-IR) interferometer from a space-based platform would bring about a transformational improvement in the spatial detail of astronomical observations that are inaccessible from ground-based observatories. This orders-of-magnitude improvement would significantly advance studies of newly formed and mature exoplanetary systems and their potential for supporting life, the earliest stages of star formation, and the evolution of galaxies up to the peak of the Universe's star formation activity and beyond. In this work, we discuss the potential for such an instrument and the science it enables in complement to current long-term plans of the global astronomical community and in preparation for the next generation of observatories. We also outline notable far-IR mission concepts, pathfinder instruments, and technology development platforms that have been studied and developed throughout the past few decades and briefly discuss key enabling technologies. 
\end{abstract}

\section{The scientific promise of a space-based far-IR interferometer} %2 Pages D:
\label{sec:science}
Approximately half of the total energy density of electromagnetic radiation emitted by stars and accreting objects in the Universe throughout cosmic time is observed in the far-infrared (far-IR) regime (loosely defined as encompassing $\sim$30 to 1000\,$\mu$m or 10 to 0.3\,THz)~\citep{Blain2002,Hill2018}. Indeed, the far-IR background~\citep{Hauser1998}, which is predominantly comprised of the light from galaxies in the early Universe~\citep{Dwek1998}, represents the second most dominant constituent of the energy content of the Universe (following the cosmic microwave background;~\citealt{Penzias1965,1990Mather}). The far-IR spectral regime offers a diversity of diagnostics; here, we highlight some notable examples. The cold and obscured early stages of star formation are revealed through continuum measurements of the dust found in sources ranging from the filamentary structures of molecular clouds to the envelopes and discs that surround newly forming stars. The far-IR provides unique access to a plethora of vibro-rotational molecular lines, including many water features, and fine structure lines from both neutral atoms and ions. Through these spectral features, the phase (i.e., cold molecular gas, warm ionised gas, etc.), dominant cooling mechanisms, and driving shocks that govern the interstellar medium (ISM) and ultimately star formation can be effectively scrutinised. Furthermore, the spectral energy distributions (SEDs) of early-stage protostars and circumstellar discs peak in the far-IR. Continuum measurements, coupled with spectroscopy of far-IR lines, provide key diagnostics for the mass of encircling envelopes, outflow dynamics, and accretion rates of these embedded targets. Importantly, the far-IR provides access to broad water-ice features that (when coupled with gas-phase water lines) allow the abundance and phase of water to be traced in planetary systems~\citep{2018Kamp}. Beyond our galaxy, this waveband is crucial in revealing the light of star-forming galaxies (in which ultraviolet light is absorbed by interstellar dust and re-radiated in the far-IR). A sensitive far-IR observatory, with sufficient spatial resolution to reach beyond the limits of source confusion (see Section~\ref{subsubsec:confusion}), would enable studies of the co-evolution of galaxies and their supermassive black holes and the accurate census of dusty star formation and the metal enrichment of galaxies across cosmic time (out to high redshift)~\citep{2026Farrah}. Furthermore, such far-IR measurements also have the potential to measure the redshifted rotational-vibrational transitions of molecular hydrogen from the environment of the Universe's first stars, which is expected to be metal-poor and highly energetic~\citep{Linz2021}. Owing to these compelling diagnostic tools, \textbf{far-IR astronomy is well positioned to significantly advance studies of new worlds/planetary systems and the search for life, the earliest stages of stellar development on to the development of circumstellar discs and their eventual dispersion, and the assembly history of galaxies~\citep{Farrah2019}.}

\begin{figure}[htbp]
    \centering
    \includegraphics[width=\linewidth]{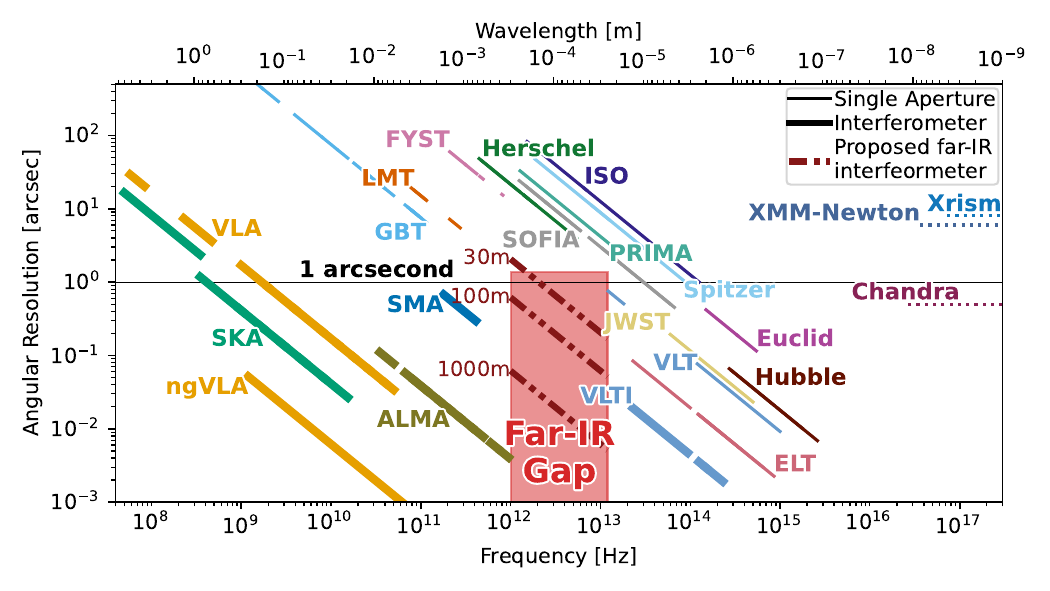}
    \caption{The finest degree of angular resolution achievable from selected major astronomy facilities (including planned observatories). The far-IR regime represents a significant gap in which sub-arcsecond observations are not available. Red lines with both a dash and dott pattern show the potential of a far-IR interferometer for a number of proposed baselines. We note that for X-ray facilities, angular resolution is not a simple matter of typical diffraction limits; for ease of comparison, their finest achievable angular resolution is instead shown as constant across their respective spectral bands.}
    \label{fig:angres}
\end{figure}

In spite of its great promise, effectively observing the Universe in the far-IR is exceptionally challenging, particularly when compared to neighboring spectral regimes. Perhaps chief among obstacles is Earth's atmosphere, which is completely opaque to the majority of the far-IR from even the best ground-based observing sites. As a consequence, the most effective far-IR observatories are space-based out of necessity. This requirement exacerbates two other significant obstacles: the need for cryogenically cooled optics and detectors, and the need for large primary apertures to address the increase in diffraction at long wavelengths (scaling as $\propto\lambda$). Any element on an observatory at temperatures higher than $\sim$5\,K represents a significant contribution in ambient radiation at these wavelengths. Even warm components with an extremely poor emissivity contribute far-IR light that greatly exceeds that from background sources (the far-IR background and zodiacal light). Furthermore, detectors capable of observing low-energy far-IR photons require sub-kelvin level temperatures to achieve the sensitivities necessary for most astrophysical targets. This challenge, coupled with that of long-wavelength diffraction has largely separated the heritage of space-based far-IR observatories into two categories: those with small cooled apertures and reduced spatial resolution (i.e., observations with limited ability to resolve spatial detail), and those with large warm apertures and reduced sensitivity (i.e., observations that are less capable of resolving faint sources). The \textit{Herschel} space observatory~\citep{Pilbratt2010} is a prime example of a large warm telescope. Having a primary mirror of diameter 3.5\,m that was only passively cooled to $\sim$85\,K (the largest monolithic mirror fielded on a space telescope, at the time of writing) represents the highest degree of spatial resolution that has been available in the far-IR (and in the sub-mm, from a space-based platform). In spite of this progress, the angular resolution of far-IR observations lags that which is readily available at neighboring wavelengths by several orders of magnitude (i.e, for the near future there is a clear ``far-IR gap''). The Probe Far-Infrared Mission for Astrophysics (PRIMA), currently under Phase B development, is positioned to provide the first far-IR telescope with a relatively large actively cooled primary mirror in the coming decade ($\sim$1.8\,m in diameter;~\citealt{Glenn2025PRIMA}). PRIMA aims to fully exploit significant advances in superconducting detectors to provide truly background-limited sensitivity (i.e., being limited only by the residual light from the solar system and the cosmic infrared background, a generational improvement in its own right). However, PRIMA does not represent any meaningful improvement to the spatial resolution of far-IR observations. The current spatial resolution gap in the far-IR is seen in Fig.~\ref{fig:angres}, which demonstrates that {\bf this is the only regime across the electromagnetic spectrum from radio to X-ray that has not yet achieved observations with sub-arcsecond angular resolution}. 

Spatial interferometry represents perhaps the most practical solution to the technical challenges that limit the spatial resolution of far-IR observatories, providing a path to both high sensitivity and a fine degree of angular resolution. To achieve sub-arcsecond resolution at say, $\lambda=100$\,$\mu$m, a 25\,m or larger telescope would be necessary, representing a considerable launch mass that must also be cooled to $\sim$4\,K and placed in an orbit about the second Lagrange point (the preferred orbit for far-IR astronomy, see~\citealt{Pilbratt2010}). For a single aperture telescope, sensitivity scales with the square of the mirror's diameter while spatial resolution only improves with diameter to first order. As such, sensitivity requirements for a given single-aperture observatory are reached before any significant improvement to spatial resolution is made. As noted by \citet{Harwit2006}, ``\textit{Like in the radio wavelength regime, large effective apertures are needed primarily to provide improved diffraction-limited images. For this purpose, sparse apertures and, in particular, interferometers, are fully adequate and economically viable.}''

\begin{table}[htb]
    \centering
    \caption{Examples of science investigations enabled by observations with a fine degree of angular resolution in the far-IR ($\sim$30 to 1000\,$\mu$m). These investigations have been noted in a series of far-IR community workshops and other works outlined in the text of the document. Many of these are published in: ~\citet{Workshop2004}, \citet{FIRI2009}, and \citet{2014Leisawitz_WS}. The approximate angular resolution requirements for each case, $\Delta\theta$, is also shown. In several cases, these represent the angular resolution goals set by the community consensus of these workshops. The maximum baselines (or approximate mirror sizes) to achieve these can be referenced against the red lines with both a dashed and dotted pattern in Fig.~\ref{fig:angres}.} %TODO Restructure following Tom's suggestions. DONE. Table consolidated and roughly grouped as ISM, Extragalactic, Planetary formation and then ordered by angular resolution within each group
    \begin{tabular}{lC{1.5cm}|lC{1.5cm}}
    Science Investigations & $\Delta\theta$ [arcsec]  & Science Investigations & $\Delta\theta$ [arcsec]   \\
    \hline\hline
    6\,m-class single aperture observatory & 8.4 & (\S~\ref{subsubsec:SF}) & \\
    30\,m-class interferometer observatory & 0.7 & ISM Star formation feedback & 1 \\
    100\,m-class interferometer observatory & 0.2 & ISM Turbulent dissipation regions & 0.7 \\
    1000-m class interferometer observatory & 0.02 & Interstellar gas evolution & 0.3--1 \\
    \cline{1-2}
    (\S~\ref{subsec:newWorlds}) & & Interstellar dust evolution & 0.3--1 \\
    Water abundance in protostellar cores & 1 & Photodissociation Regions & 0.1 \\
    Debris disks (resonant structures) & $<1$ & Turbulent ISM & 0.1 \\
    HD/CO-dark gas & 0.01--1 & Asymptotic giant branch shells & 0.01--0.1 \\
    Solar system science & 0.1 & (\S~\ref{subsec:galaxies}) & \\
    Water tracing infall & $<0.1$ & Extragalactic giant molecular clouds & 1 \\
    Protostellar outflows & 0.05 & AGN--star formation coevolution & $<1$ \\
    Protoplanetary discs: water & 0.02--0.1 & Galaxies (incl. ULIRGs): Dust & $<0.05$ \\
    Planetary formation in nearest protostars & 0.01 & Galaxies (incl. ULIRGs): Gas & $<0.05$ \\
    Protoplanetary discs: dust & $<0.01$ & AGN line diagnostics & $<0.05$ \\
    Debris discs: planetary systems & $<0.01$ & Deep field targets & 0.02 \\
    Giant planets in protoplanetary discs & 0.001 & Merging AGN & $<0.02$ \\
    Giant planets in debris discs & 0.001 & Luminosity function in clusters & 0.01 \\
    \end{tabular}
    \label{tab:sciencecases}
\end{table}

{\bf We propose that a space-based far-IR interferometer leveraging the double-Fourier technique be prioritised in the coming decade to provide the transformational leap in spatial resolution necessary to close the far-IR gap}. Indeed, we re-issue the sentiment set out in the roadmaps of previous far-IR community workshops and white papers that, following a large, cooled single-aperture facility (i.e., PRIMA, which NASA recently selected for Phase-B study), a mission that couples the sensitivity of these with the spatial resolving capabilities of an interferometer be the next step for far-IR observing~(\citealt{Leisawitz2004,2013Leisawitz} and references therein; \citealt{Sauvage2013,EUfarIRRoadmap}). The interferometer will build on preparatory work on the Space Infrared Telescope for Cosmology and Astrophysics (SPICA), which is no longer in development, and PRIMA, as well as decades of interferometry mission concept studies, pathfinder missions, and technology development platforms (many of which are outlined within Section~\ref{sec:concepts}), that have identified the double-Fourier technique~\citep{1988Mariotti,Itoh86} as the most promising pathway to space-based interferometry in the far-IR.

The science cases that are addressed by the generational leap in spatial resolution that a far-IR interferometer provides range over almost all areas of astrophysics and cosmology. In particular, a series of far-IR community workshops held in College Park, Maryland (2002), Madrid (2004), Leiden (2005), Obergurgl (2006), and Pasadena (2014) noted the need for sub-arcsecond angular resolution observations in the far-IR to advance fields of study ranging from studies of gas and dust in circumstellar discs, studying interstellar medium (ISM) turbulence and star formation feedback, and on to ultra-luminous IR galaxies (ULIRGs) and cosmological deep fields~\citep{Workshop2004,FIRI2009}. In Table~\ref{tab:sciencecases} we note a few examples of the science that is unlocked by sub-arcsecond angular resolution in the far-IR as reported from the discussions at these workshops and also from additional works referenced throughout this white paper~\citep{ESA2050FIRinter,Leisawitz2004,Rinehart2016,Rinehart2019}. 

In the following section, we detail some of the most compelling science gaps that a far-IR interferometer directly addresses. These are framed in the context of open questions that have been highlighted in various long-range planning documents from various astronomy communities, in particular the UK Research and Innovation Science and Technology Facilities Council Science Challenges in Frontier Physics~\citeyearpar{STFC2019}, the National (USA) Academies of Sciences, Engineering, and Medicine ``Pathways to Discovery in Astronomy and Astrophysics for the 2020s''~\citeyearpar{Astro2020}, The National (USA) Aeronautics and Space Administration (NASA) Astrophysics Roadmap ``Enduring Quests, Daring Visions''~\citeyearpar{Roadmap2014}, the Astronet Science Vision and Infrastructure Roadmap 2022--2035~(EU, \citeyear{Astronet2022}), and the Canadian Astronomy Long Range Plan 2020--2030~\citeyearpar{LRP2020}, hereafter referred to as Frontier2019, Astro2020, Roadmap2013, Astronet2022, and LRP2020, respectively.

\subsection{New stars and new worlds}
\label{subsec:newWorlds}
%hot spots in the disc, temperature differences at peak of SED
%Water ice features, peak emission of debris disks. See strcture over ALMA ~A little less: Noted in Brenda's paper
%confusion limit
%resolving individual high-redshift galaxies at 100m
%Distinguish between SF and central AGN through spectral lines (ALMA can only do high Z). Struggle with line sensitivty. Dispersion not an issue for distant galaxies, these are point sources.
%Trade study on what to add to SPICE. 50m
%Boom good up to about 100m
%Lorenzini tethered strcuture.
\epigraph{What processes govern how planetary systems form and evolve?\\ What are the conditions for life and how widespread are they?\\ How diverse are exoplanets and is our earth typical?}{--- Frontier2019 (UK)}
\vspace{-1cm}
\epigraph{How do habitable environments arise and evolve within the context of their planetary
systems?\\ How can signs of habitable life be identified and interpreted in the context of their
planetary environments? \\ How does gas flow from parsec scales down to protostars and disks? \\ Is planet formation fast or slow?}{--- Astro2020 (USA) Panels on ``Exoplanets, Astrobiology, and the Solar System'' and ``The Interstellar Medium and Star and Planet Formation''}
\vspace{-1cm}
\epigraph{Understanding the assembly of planets, stars and galaxies requires information across the entire electro-magnetic spectrum, with a particular need in the next decades for new far-infrared and UV space telescopes.\\
The continued characterisation of exoplanet systems will require direct-imaging capabilities, and FIR/optical/UV spectroscopic capabilities to characterise atmospheres, or even infrared space-based interferometry.}{--- Astronet2022 (EU)}
\vspace{-1cm}
\epigraph{The key first step to understanding how planets form and evolve is to measure the composition of exoplanets and their atmospheres ... These studies must be accompanied by observations of the ``protoplanetary'' disks of gas and dust that surround newly-formed stars.}{--- LRP2022 (Canada)}

\begin{figure}[b]
    \centering
    \includegraphics[width=\linewidth]{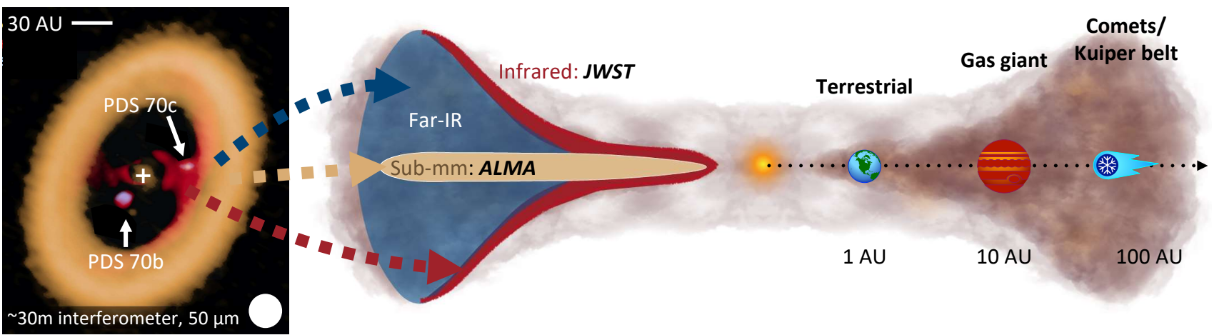}
    \caption{Illustration showing the structure of a protoplanetary disc. While ALMA and JWST are capable of probing the disc's midplane and surface, respectively, far-IR observations reveal the bulk of the disc volume. A far-IR interferometer would be capable of a tight enough beam (inset circle in the left panel) to map the disc's mass distribution and water reservoir, ultimately tracing the differences in the origins of lifeless and potentially habitable planets. Credits: ALMA (ESO/NRAO/NAOJ) \& A. Isella/ESO; M.K. McClure.}
    \label{fig:disc}
\end{figure}

\subsubsection{The trail of water}
One of the primary motivations for space-based far-IR observation is to measure transitions of the cold water molecule that cannot be effectively observed from the ground. Water is a key tracer of star formation processes and plays a crucial role in planetary formation and understanding the composition of protoplanetary systems. Furthermore, and perhaps most evidently, water is critical for the emergence of life and evolution as we know it~\citep{2017Pearce}. Indeed, water is a primary constituent in even the most primitive cellular life on Earth. The far-IR is uniquely suited for tracing the quantity and phase of water in protoplanetary systems, having both low-lying (i.e., cold) water-ice spectral features and low-lying water-gas features. With its unprecedented level of spatial detail and its imaging spectroscopic capabilities, a far-IR interferometer will push frontiers of discovery in the search for extrasolar life as per the priorities set in the NASA 2022 Strategic Plan (SP2022) and the European Space Agency's (ESA) Cosmic Vision 2015-2025. A far-IR interferometer could directly map the distribution and phase of water in newly forming planetary systems, tracing the processes that make the difference between a potentially habitable world and a lifeless one.

Of particular note, the 45 and 63\,$\mu$m water-ice features found in the far-IR provide a critical diagnostic in planetary formation processes. Icy mantles on dust grains increase how efficiently they stick together and accumulate into planetesimals. Furthermore, water serves as the dominant oxygen carrier in discs, being particularly important in the chemistry of discs, with the ratio of carbon and oxygen (C/O) providing an important link between the composition of discs and the atmospheres of protoplanets~\citep{2011Oberg,2017Cridland,2019Pontoppidan}. The water ``snowline'' in discs is the radial distance at which the temperature drops sufficiently for gas-phase water to transition to the solid phase. Complementing measurements of snowlines of other species with JWST~\citep{2023Sturm}, the water-snow line is useful in tracing different C/O ratios in the gas and in icy dust mantles at different disc radii~\citep{Piso2015}. We note that measuring the water-ice snowline in protoplanetary systems, albeit in a spatially-unresolved fashion, is a science priority for the PRIMA-FIRESS instrument. Measurements at other wavelengths are limited to highly inclined discs to measure features from water-ice in absorption, which is then complicated by radiative transfer effects and contamination from scattering~\citep{2025PRIMABook2, 2021Ballering, 2023Sturm}. Direct imaging spectroscopy of protoplanetary systems with the fine degree of angular resolution and spectral capabilities of a far-IR double-Fourier interferometer would greatly improve our understanding of the distribution of the water snowline in these systems to predict and constrain the types of planet and comet compositions that result from different parts of the disc~\citep{Leisawitz2023,Piso2015,ESA2050FIRinter,Aikawa2024, 2011Hogerheijde}. 

%Of particular note, the broad 45 and 63\,$\mu$m water-ice features found in the far-IR provide a critical diagnostic in planetary formation processes. Icy mantles on dust grains increase how efficiently they stick together and accumulate into planetesimals. Coupled with measurements of other volatile species with far-IR ice features (CO, CO$_2$, and NH$_3$), understanding the spatial distribution of ices (i.e., snowlines) would allow us to constrain where comets and planets of different compositions form in protoplanetary discs~\citep{ESA2050FIRinter, Aikawa2024, 2011Hogerheijde}. We note that measurements of these, albeit in a spatially-unresolved fashion, is a science priority for the PRIMA-FIRESS instrument, as measurements of these species at other wavelengths are limited to highly inclined discs to measure these features in absorption, which is then complicated by radiative transfer effects and contamination from scattering~\citep{2025PRIMABook2, 2021Ballering, 2023Sturm}. Additionally, the ratio of carbon and oxygen (C/O) has been shown to provide an important link between the composition of discs and the atmospheres of protoplanets, with water playing a key role as the dominant oxygen-carrier in discs~\citep{2011Oberg,2017Cridland,2019Pontoppidan}.

In these efforts, we echo \citet{ESA2050FIRinter} in noting the shortcomings of the single-aperture facilities. Such observatories would be limited to measurements of water abundance in only a spatially integrated fashion or would be subject to model-based biases. Measurements from Herschel-HIFI over a sample of T Tauri discs suggest the oxygen abundance is not homogeneous across the entire disc and that much of the water is frozen onto dust grain surfaces~\citep{2017Du}. A far-IR space interferometer is necessary to build a complete picture of the structure and phase of water in newly forming planetary systems. Indeed, such a facility would separate water-gas emission from inside the H$_2$O snow line from emission arising from layers that are further out~\citep{Leisawitz2023}. Additionally, the broad spectral band of the double-Fourier technique is well matched to measure ice features which are too broad to be effectively measured through heterodyne spectroscopy (having widths of order $\Delta\lambda\sim30$\,$\mu$m). Understanding the distribution of water in protoplanetary disks may provide key insights into the difference between habitable worlds and lifeless planets. Through spatially resolved measurements of water abundance and phase, a far-IR interferometer directly addresses the priority of ASTRO2020 to trace the incorporation of water and atmospheric volatiles into the overall constitution of planets and their atmospheres/potential biospheres~\citep{2012Marty}.

\subsubsection{CO-Dark molecular hydrogen}
The majority of molecular gas throughout the ISM and within circumstellar disks of new planetary systems consists of H$_2$ molecules, which do not possess a permanent dipole moment. As a consequence, the bulk of the molecular mass of these systems is not directly measured and proxy molecules (most commonly CO and N$-2$H$^+$) are needed~\citep{2025Zhang}. While the low-lying rotational transitions of CO are accessible with a fine degree of spatial resolution from ground-based sub-millimetre (sub-mm) facilities such as ALMA, these are subject to significant systematic uncertainties that make measurements difficult to interpret. Indeed, within protoplanetary disks, CO may be frozen onto dust grains (and thus non-emissive), photodissociated within the disc atmosphere, or removed from the gas phase by chemical processes, and ultimately the precise ratio of CO to H$_2$ is not conclusively known~\citep{2017Molyarova,2018Schwarz}. Furthermore, CO-dark H$_2$ is a problem that extends beyond protoplanetary systems, with ``CO-dark'' molecular gas possibly contributing 20--80\,\% of the molecular mass in our Galaxy~\citep{2014Pineda,2015Grenier,Smith2014}.

Molecular hydrogen's close deuterated relative, HD, has been demonstrated to be a robust tracer of H$_2$ and has low-level rotational lines at 56 and 112\,$\mu$m. As these transitions are not readily accessible from the ground, they have been noted as an important science goal for future far-IR missions to provide an independent mass tracer for circumstellar gas with a relatively low uncertainty~\citep{2013Bergin,2016McClure,2017Trapman}. A space-based far-IR interferometer would be capable of unbiased measurements to clearly resolve the mass distribution of gas in circumstellar disks.

In complement to measurements of HD, IR fine structure lines ([CII], [OI], and [CI]) and the ground-state rotational transitions of light hydrides (CH, HF, CH$^+$, and OH) throughout the far-IR have been shown to be effective tracers of CO-dark H$_2$ in the ISM~\citep{2013Pineda, 2018Rugel, 2019Jacob}. Measuring these features and resolving their velocity structures with low-resolution spectroscopy is within the science goals of the PRIMA mission to trace different ISM components and develop a galactic-wide ISM tomography~\citep{2025PRIMABook2}. A fine degree of spatial resolution would complement these by spatially resolving the boundaries of transition regions and the turbulent heat dissipation that dominates the equation of state for ISM gases in three dimensions, leading to a complete picture of the ISM's evolution towards star formation~\citep{ESA2050FIRinter}. Extragalactically, the power to observe these lines at high spatial resolution has never existed in the past (with ALMA only able to access them at high-$z$). A far-IR interferometer will enable tracing these on the scales of giant molecular clouds in external galaxies, extending the range of environments where we can study star-formation science in detail, opening up regimes which simply do not exist in our own Milky Way. 

\subsubsection{Probing the architecture and evolution of planetary systems via debris discs}

Every star, from the ages of 10s of millions to billions of years old, is likely accompanied at some time in its evolution by a debris disc composed of planetesimal belts and their collisional products of dust and gas. The ubiquity of these discs and resonant substructures within them provides a window into planetary architectures over the epochs of terrestrial planet formation, planetary migration, and late-age mature planetary systems, i.e., potential Solar System analogues \citep{Lissauer1993, Raymond2011Debris, Wyatt2003, Wyatt2008, 2026Marino}. The presence of substructure and asymmetries is prevalent in observations capable of resolving them (e.g. \citealt{Lovell2025,Crotts2024,2025Matra}). Such substructures in dust and gas emission can reveal planets, including constraints on their mass and orbits, as well as provide compositional insight into early planetary evolution (such as the delivery of volatiles to nascent planets) \citep{Pawellek2014}. Debris discs are detectable because collisional evolution within them produces dust and gas debris with a very favorable cross-section-to-mass ratio, particularly compared with relatively small planets, such as rocky terrestrial planets or ice giants on very long orbits. Ice giants provide unique insight into the earliest phases of planetary systems despite being particularly challenging to detect due to the long timescales of their orbits. 

\begin{figure}
    \centering
    \includegraphics[width=0.6\linewidth]{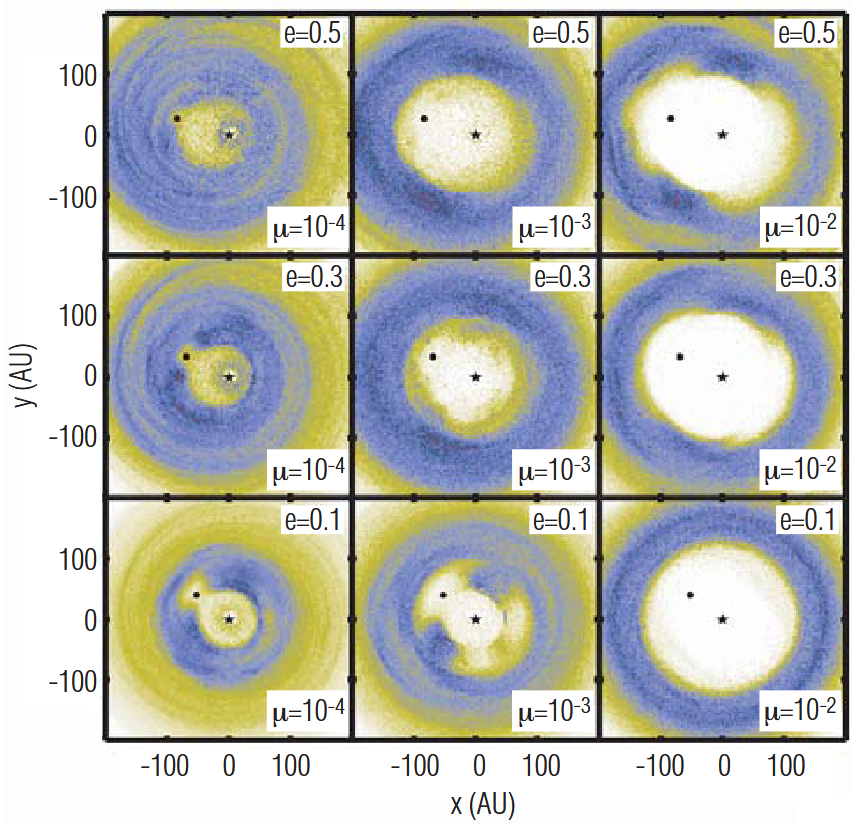}
    \caption{Models of debris disc emission at 100\,$\mu$m showing resonant structures that would be uncovered by a far-IR interferometer. The obscured planet ($\bullet$) has a mass that is a fraction, $\mu$, of the mass of the star $\star$ and an orbit with eccentricity $e$. Figure provided by \citet{SPIRITPropBook}.}
    \label{fig:debrisdisc}
\end{figure}

Debris discs are inherently fainter than their protoplanetary predecessors, with measured dust fractional luminosities from 10$^{-3}$ to a few 10$^{-6}$, still a factor of several above that of the Solar System~\citep{Hughes2018}. Dust temperatures place their peak emission in the mid- (warm components) to far- (cold components) infrared; but due to historically limited spatial resolution in the far-IR, ground-based interferometric observations at submillimetre wavelengths (where emission is at least two orders of magnitude fainter) have been relied upon to detect the types of substructures seen in Fig.~\ref{fig:debrisdisc} and thereby infer the architecture of planetesimal belts. This requires significant time investment to obtain moderate-sized population studies~\citep{Marshall2021}. Such studies are therefore limited to relatively bright (typically nearby) disks and hence relatively few known debris discs have even been observed with ALMA. Many more discs would be accessible to sub-arcsecond imaging in the far-infrared. Despite \textit{Herschel}'s unprecedented mirror size (3.5 metres in diameter), it was only able to resolve the largest of debris discs~\citep{Marshall2021}. Even a modest far-IR interferometer (baselines of 10s to 100m) would be capable of resolving structures such as rings and gaps in hundreds more debris discs. Notably, the double-Fourier technique provides multi-octave SED measurements of the dust continuum from these sources, yielding mass sensitivity to both warm and cold dust components. 

In addition to detecting warm components, a space interferometer reaching wavelengths as short as 25 $\mu$m would also be sensitive to young exoplanets shaping their debris discs. With MIRI on James Webb, \citet{Lagrange2025} directly detect a planet around the young star TWA 7, the existence of which was inferred from disc structure observed in scattered light. A mid- to far-IR interferometer therefore would have sensitivity to both substructures in debris discs as well as the exoplanets that sculpt them, testing models of planet-disc interactions as never before.  

\subsubsection{Tracing the earliest stages of star formation}
\label{subsubsec:SF}
The far-IR allows access to a multitude of molecular rotational transitions (including water, carbon monoxide, and other light hydrides and carboniferous molecules;~\citealt{Tielens2005}) and atomic fine structure lines ([CII], [NII], [CI], [OI];~\citealt{2013Pineda,Goldsmith2015}) that, together, provide key tracers of the interstellar medium (ISM) and its role in regulating star formation and overall galaxy evolution~\citep{Shu1987,ramos2023}. Through these, cooling processes, ionisation regions, and shocks that ultimately govern star formation can be traced. It must also be noted that in all of these studies, a fine degree of angular resolution and sensitivity would also allow the first extension of many of these to nearby galaxies, linking the Milky Way to the wider galaxy population. Furthermore, emission corresponding to the earliest stages of protostar development peaks in the far-IR, making continuum measurements in this spectral band crucial in studying these.

In recent years, the sub-arcsecond resolution provided by ALMA's interferometry has provided significant insights into the origins of the initial mass function (IMF, the relative number of stars born at different masses) through the ALMA-IMF Large Program~\citep{2022Motte,Pouteau2022,Armante2024,Louvet2024,ValeilleManet2025}. The IMF is important to most areas of cosmology and stellar physics with several open questions under debate (\citealt{Lee2020,Krumholz2014,2022Motte}, and references therein). ALMA-IMF aims to provide a detailed scrutiny and study of statistical samples of massive protoclusters to test these observationally. Similarly, sub-arcsecond resolution mapping programs are providing significant advances in studies of high-mass star formation through The ALMA Survey of 70\,$\mu$m dark High-mass clumps in Early Stages~\citep[ASHES;][]{Sanhueza2019} and the role of molecular clouds and star-formation complexes in driving Galactic evolution through the ALMA central molecular zone (CMZ) Exploration Survey~\citep[ACES;][]{Longmore2026}. In spite of this great progress in all of these, ALMA's spectral coverage measures only the tail end of the SEDs from protostellar cores. Thus, the majority of light from these sources has not been observed with a similar level of spatial detail, owing to the poor angular resolution available in the far-IR. A far-IR interferometer (bridging between ALMA and JWST) that is suited to large-scale mapping is well-positioned to complement these surveys, providing detailed measurements of fainter sources and structures that will significantly advance nearly all areas of study in star-formation.

%Through the fine degree of spatial resolution provided by ALMA, several key questions concerning the role of molecular clouds and star-formation complexes in driving Galactic evolution are being addressed through large surveys such as the ALMA Central Molecular Zone Exploration Survey that can connect large-scale gas flows down to the sub-parsec dense structures of the host star-forming cores~\citep{Longmore2026}. In spite of this great progress, the majority of the SED of these regions is not observed with a similar level of detail owing to the lack of angular resolution in far-IR observations to bridge between ALMA and JWST. Indeed, the SEDs corresponding to the earliest stages of protostar development peak within the far-IR and many of the spectral transitions of these key gas-dynamic tracing species are optically thin within the far-IR, allowing us to discover fainter sources, better constrain mass and gas reservoirs, and peer deeper within these structures~\citep{Shu1987,ramos2023}.

In spite of the wealth of ISM diagnostics within the far-IR band, much of the details of ISM structure is not spatially resolved in the observations of single-aperture facilities. Velocity-resolved measurements of massive star forming regions in the Milky Way indicated spatial scales corresponding to $\lesssim1$\,arcsecond in nearby regions of the Galaxy~\citep{Linz2021}. Indeed, regions of the ISM where turbulence plays an important role in energy dissipation also represent structures on the sub-arcsecond scale~\citep{2009Godard}. Measurements of these structures within the ISM through spatially resolved maps of far-IR spectral features are crucial in developing a more complete picture of the collapse of molecular clouds into stars, settling whether this is always driven through the virial parameter (i.e., predominantly-self gravitation) or external factors~\citep{Krumholz2025,VazquezSemadeni2026,Lin2025}. Even a modest far-IR interferometer (maximum baselines of order $\sim30$\,m) would provide significant advances toward a complete picture of the structure of ISM and the star formation processes within.

\subsection{Galaxies out to cosmic noon and beyond} %looking for a figure to sell here
\label{subsec:galaxies}

\epigraph{Understanding the evolution of stars and galaxies from the time of their formation, through cosmic time, up to the present day is one of the major challenges in modern astrophysics.}{--- Frontier2019 (UK) Challenge A5}
\vspace{-1cm}
\epigraph{How did the intergalactic medium and the first sources of radiation evolve from cosmic
dawn through the epoch of reionization?\\ How do the histories of galaxies ... shape their observable
properties? \\ How do gas, metals, and dust flow into, through, and out of galaxies?}{--- Astro2020 (USA) Panel on Galaxies}
\vspace{-1cm}
\epigraph{To fully capitalise on the discoveries that JWST will enable, we will need the capabilities of ... new far-infrared/submillimeter facilities to characterise the interstellar medium, stellar populations and black holes of the first galaxies, and their impact on the intergalactic medium.}{--- Astronet2022 (EU)}
\vspace{-1cm}
\epigraph{How did matter interact
with gravity and other forces to form the first stars and galaxies? \\ How have galaxies changed over time? \\ How has the chemical composition of the Universe evolved? \\ What turns the process of star formation on and off in a galaxy? \\ Why is the birth rate of new stars today so much lower than it was in the distant past?}{--- LRP2020 (Canada)}

Measurements of the far-IR sky with a fine degree of angular resolution will play a vital role in understanding the picture of evolution and assembly history of galaxies. Measurements from 40 to 400\,$\mu$m reveal dusty galaxies and embedded active galactic nuclei (AGN) over a wide range of redshifts. Simulations of IR-luminous galaxies have shown that far-IR observations are key at reducing biases in luminosity measurements achieved through SED fitting, particularly at high redshift. In particular, far-IR spectroscopy was shown to reduce systematic bias and random error in the recovery of total and starburst luminosity over measurements with mid-IR spectroscopy alone by at least a factor of 4, and substantially enhances the recovery of AGN and host luminosity in conjunction with other instruments across the spectrum. In this luminosity recovery, angular resolution plays a key role in reducing systematics, with arcsecond to sub-arcsecond resolutions allowing starburst luminosities to be recovered with almost no systematic bias~\citep{2026Farrah}. Furthermore, \citet{2024Mateo} (and references therein) have noted that sub-arcsecond-resolution far-IR observations are needed to reveal faint sources and overcome the biases intrinsic in measurements with 2-m class telescopes. Indeed, their simulation work demonstrates that spatially integrated observations of these sources are subject to significant line blending and both over- and under-estimations of line emission. Ultimately, a far-IR interferometer that is capable of resolving individual galaxies is a necessary step to properly address Astro2020's goals concerning the formation and evolution history of galaxies. The authors' work suggests that at 350\,$\mu$m a 1000\,hour 0.5\,deg$^2$ survey with a far-infrared interferometer similar to the connected structure mission concepts outlined in Section~\ref{sec:concepts} would detect as many as $\sim$61\,000 star-forming galaxies out to $z\leq 6$.

Through its fine degree of spatial resolution, ALMA has been revolutionizing the use of fine-structure lines to study high-redshift galaxies. Reaching to finer spatial resolution in the far-IR will open up similar capabilities to study the environments of galaxies during the peak of cosmic star formation at redshifts $z\sim1-3$ (and, indeed, at lower redshifts where galaxy quenching can also be probed). With the coming era of spatially integrated measurements of the extragalactic sky with PRIMA, a massive leap in spatial resolution capabilities in the far-IR will be needed to measure the apparent sizes of galaxies in this range, which are anticipated to be as small as 1-2\,arcsec~\citep{Sahni2000}. Importantly, the spatial resolution enabled by an interferometer would enable a definitive separation of each galaxy's emission into components attributable to AGN and star-formation activity, significantly advancing studies of the co-evolution of AGN and their host galaxies and how this relates to the cosmic peak in star formation activity.

As has been noted in previous sections, the spatial resolution an interferometer provides is crucial in opening up detailed studies of the ISM and its structure in other galaxies with environments different to our own. Indeed, just as single aperture facilities have revolutionised our understanding of the ISM and star formation processes in our Galaxy (and continue to do so), a far-IR interferometer capable of $\lesssim$1\,arcsecond angular resolution would provide the same level of insights in nearby galaxies~\citep{Linz2021,Leroy2021}. In this, the far-IR is crucial in complement to longer-wavelength interferometric measurements, providing access to fine structure features that better trace the gas reservoirs of giant molecular clouds (GMCs) and trace the structure of the ionised ISM, which has been shown to extend beyond neutral atomic and molecular emission in GMCs~\citep{Linz2021,2017Bisbas}.

\subsubsection{The confusion noise limit} %http://herschel.esac.esa.int/Docs/Herschel/html/ch04s03.html
\begin{figure}[htb]
    \centering
    \includegraphics[width=0.7\linewidth]{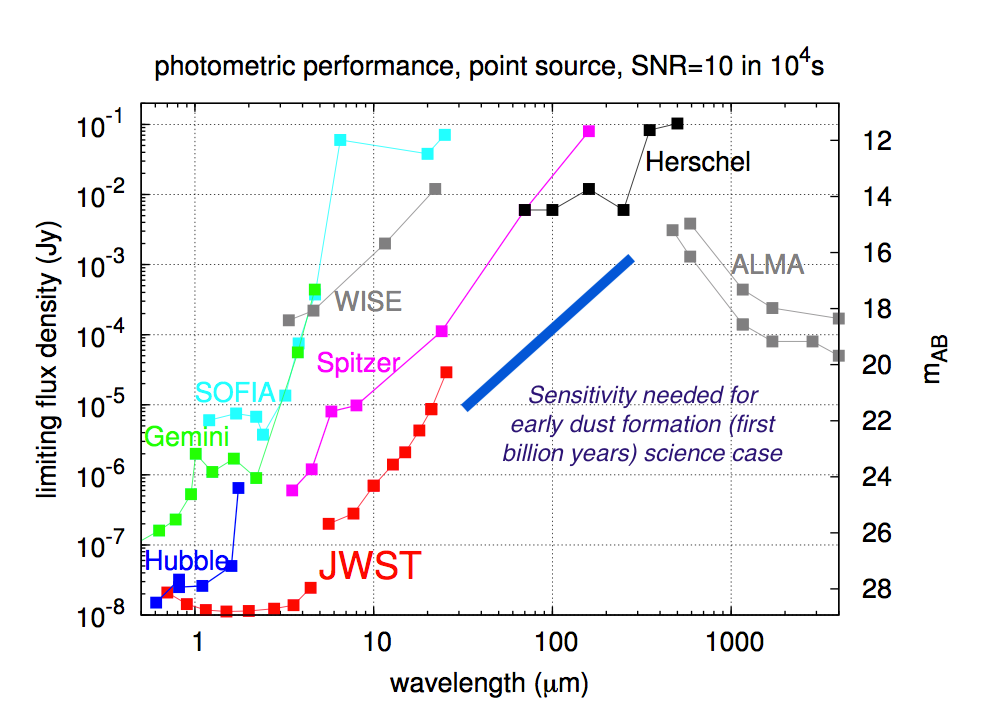}
    \caption{The sensitivity of a far-IR observatory required to study early dust formation/evolution. This level is currently beyond any existing or planned facility. Furthermore, it demonstrates that despite Spitzer's cryogenic primary aperture (during its cold-phase mission) and sensitive detectors~\citep{Moseley2002,HIROMOTO1989}, its sensitivity to such extragalactic targets was significantly hindered by confusion noise as a consequence of its poor spatial resolution. Adapted from NASA/STScl via Renske Smit.}
    \label{fig:dustSensitivity}
\end{figure}

\label{subsubsec:confusion}
The presence of spatially unresolved sources in a sky map results in a fundamental limit to sensitivity in the form of confusion noise that cannot be improved by increasing integration time. Furthermore, it becomes difficult to determine whether the origin of far-IR emission is from an extragalactic target itself, nearby sources at the same redshift, or unrelated sources within a given beam, and not all of this can be correlated with spatially resolved measurements in different wavebands~\citep{Hatziminaoglou2025}. These challenges are particularly troublesome for extragalactic surveys seeking to image a high density of faint sources. With the fine degree of spatial resolution from \textit{Herschel}'s 3.5\,m mirror and its relatively high degree of instrument noise, this confusion noise limit could be reached after 3--5 repetitions of a particular map made with its Spectral and Photometric Imaging Receiver (SPIRE) photometer~\citep{2010Nguyen,Rigby2011}. While PRIMA is well-positioned to provide a massive leap in the sensitivity of far-IR observing capabilities with a sizable aperture (${\sim}2$\,m), \citet{2024Bethermin} note that the advantage of this sensitivity cannot be fully exploited at longer wavelengths as a consequence of its limited angular resolution. Instead, PRIMA's instruments must rely on deblending methods which have been simulated to improve over the sensitivity of the confusion limit by a factor of 2--3 using priors from measurements in its shorter wavelength bands~\citep{Donnellan2024}. However, such measurements only provide a modest improvement that will be subject to the biases inherent in these methods, some of which we have previously alluded to~\citep{2024Mateo}. The sub-arcsecond resolution provided by a far-infrared interferometer is necessary to measure sources beyond the current confusion noise limit (see Fig.~\ref{fig:dustSensitivity}).

Notably, reaching beyond the confusion noise limit has a particular importance in revealing faint galaxies at high redshift that are beyond the reaches of JWST bands. Indeed, the dust-obscured star formation of typical galaxies at $z>4$ is only viewed at the $\mu$Jy level, leading to notable biases toward the most IR-luminous and/or gravitationally-lensed galaxies in formation rate density studies~\citep{2024Mateo}. Furthermore, the spectral signatures of polycyclic aromatic hydrocarbon (PAH) features from these galaxies provide key insights into the role of dust evolution in the earliest galaxies.

\begin{figure}[h]
    \centering
    \includegraphics[width=0.35\linewidth]{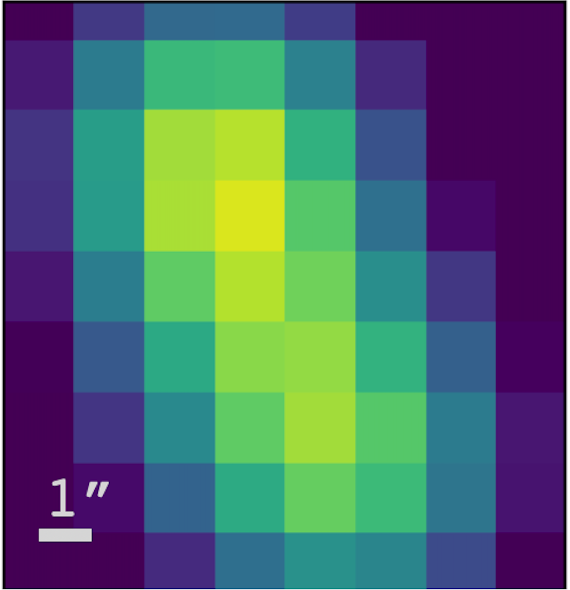}
    \includegraphics[width=0.35\linewidth]{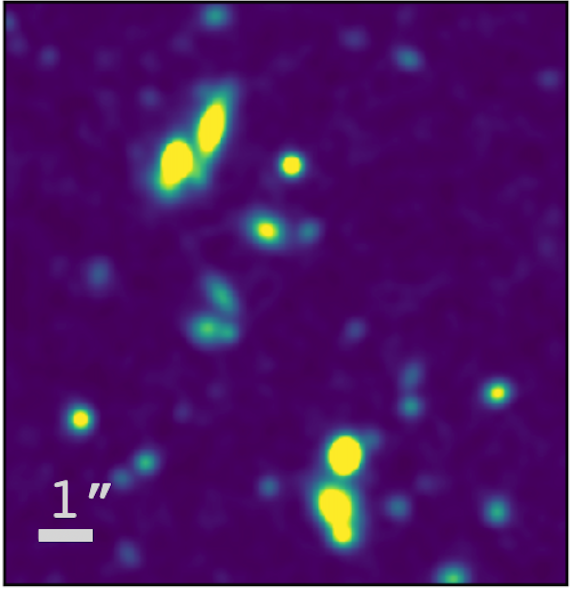}

    \includegraphics[width=0.49\linewidth,trim={0mm 7mm 0mm 7mm},clip]{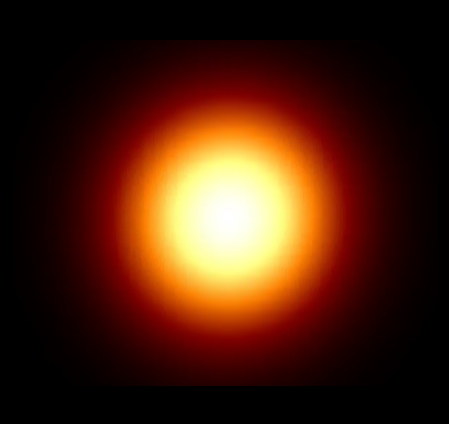}
    \includegraphics[width=0.49\linewidth,trim={0mm 7mm 0mm 7mm},clip]{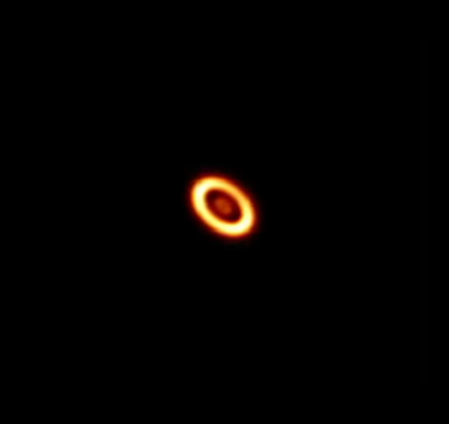}

    \includegraphics[width=0.49\linewidth]{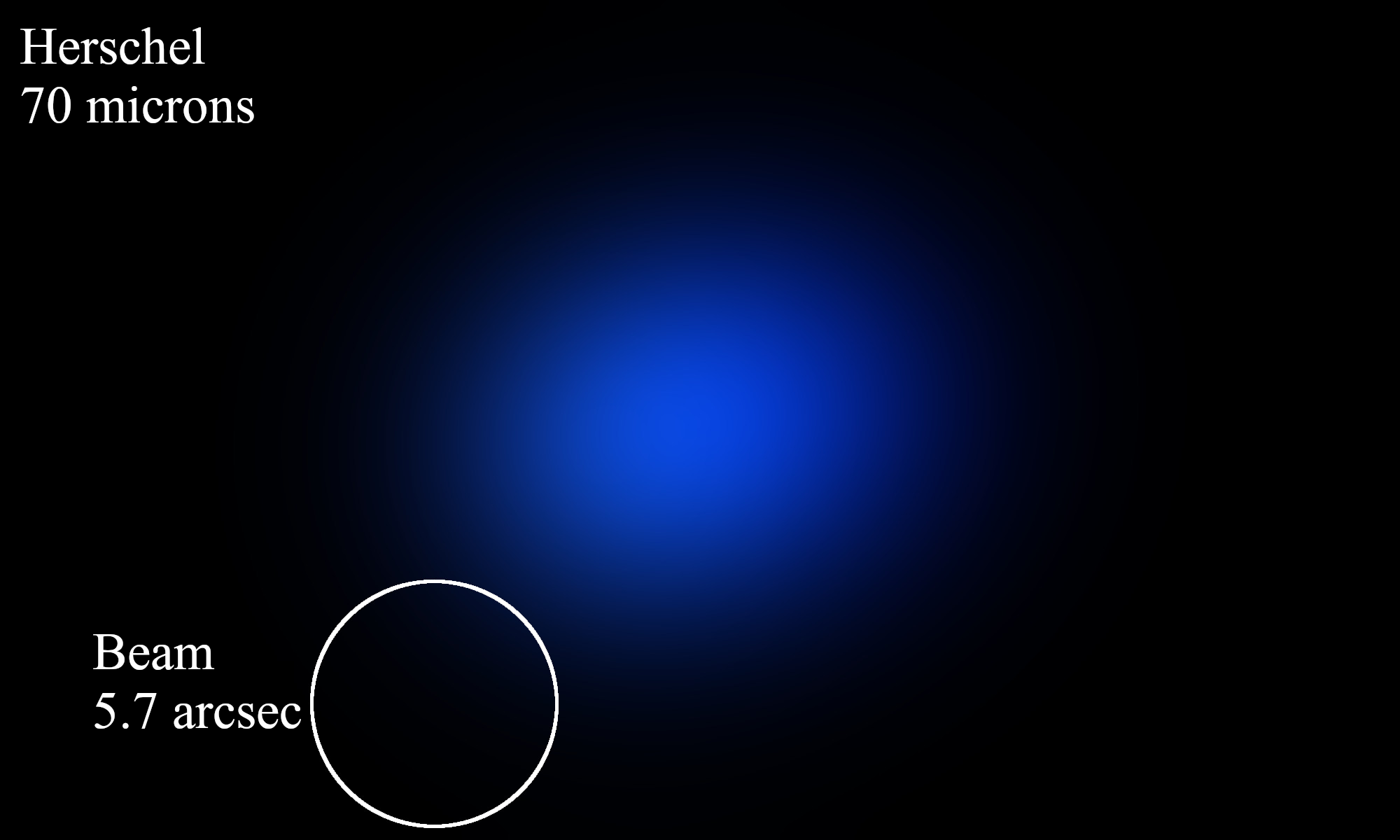}
    \includegraphics[width=0.49\linewidth]{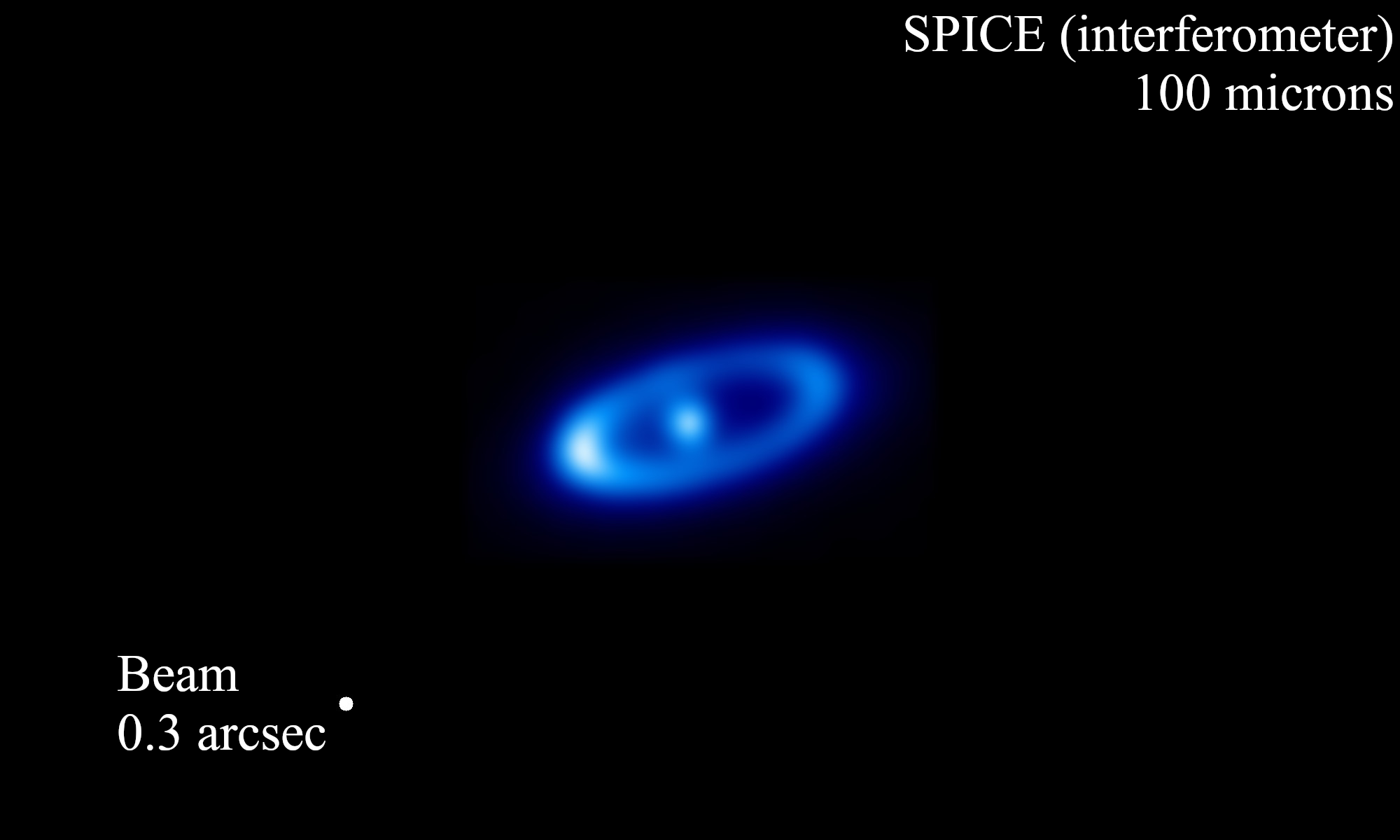}
    \caption{A comparison of the level of spatial detail of a single aperture facility (mirror diameter of $\sim$3.5\,m, left) and that of a modest interferometer (maximum baseline of $\sim$30\,m, right). Note that the cover image is a subset of these. Top Row: A simulated extragalactic field, the scale marking denotes a 1\,arcsecond extent. A single aperture facility with a mirror of a similar size to \textit{Herschel} is unable to resolve individual sources and assign separate flux contributions to these (\S~\ref{subsec:galaxies}). Middle row: A simulated view of the LkCA15 protoplanetary disc at 100\,$\mu$m. A far-IR interferometer would be capable of mapping the structure and phase of water and other constituents, providing insights into how these are incorporated into exoplanetary atmospheres (\S~\ref{subsec:newWorlds}). Bottom row: A simulated observation of the Fomalhaut debris disk as if it were at a distance of 100\,parsecs, the beam size corresponding to each facility is shown by the circle in the bottom left. A far-IR interferometer would measure the morphology of such discs revealing density enhancements and resonant structures that reveal the presence of planets. Images courtesy of the SPIRIT/SPICE instrument and simulation team.}
    \label{fig:scienceSummary}
\end{figure}

Furthermore, by fully resolving extragalactic sources, the correct attribution of IR flux to individual components in the field with counterparts at other wavelengths becomes possible. As an example, we can consider an extragalactic field with an AGN. A single aperture telescope would blend the IR flux from other galaxies within the the large beam, representing a significant contribution of flux that cannot be separated~\citep{2024Mateo}. A far-IR interferometer would be able to directly measure the SEDs of individual galaxies and provide clear flux assignment to each, and in nearby galaxies, separate out far-IR contributions from the central AGN and those originating from the rest of the galaxy. This is well-demonstrated in the top row of Fig.~\ref{fig:scienceSummary}. PRIMA is positioned to vastly open up the far-IR sky with much broader and deeper surveys than \textit{Herschel} that will uncover new sources and enable new science. A far-IR interferometer will be necessary to provide targeted observations that resolve the distribution and structure of these blended sources.

\subsection{Discovery space and greater vision}
\epigraph{Are we alone? How did we get here? How does the universe work?}{--- Roadmap2013 (USA).}
\vspace{-1cm}
\epigraph{Astronomy captures the imagination,
enriches our culture, and pushes the
boundaries of technology.}{--- LRP2020 (Canada).}
%PRIMA is positioned to vastly open up the far-IR sky with much broader and deeper surveys than \textit{Herschel} that will uncover new sources and enable new science. A far-IR interferometer will be necessary to provide targeted observations that resolve the structure and spatial distribution of these

Perhaps the most compelling reason to develop a far-IR interferometer is the vast discovery space that an orders-of-magnitude improvement in spatial detail affords. While some of the scientific endeavors discussed throughout this section can be addressed, in part, through pairing high spectral resolution, which can measure structure along the line of sight, with robust modelling, a generational leap in spatial resolution is the only way to provide direct measurements that will substantially advance \textit{all} of these areas. Furthermore, the orders-of-magnitude improvement in angular resolution a far-IR interferometer enables will simultaneously open up a vast discovery space. Indeed, Astro2020 notes that our understanding of the Universe is transformed by observing in new ways that uncover previously hidden phenomena (noting a much finer angular resolution as one of these). Here, we note the past examples in the successes of both IRAS and ALMA in revealing unseen phenomena. The first unfettered far-IR all-sky access provided by IRAS revealed thousands of massive star-forming Ultra Luminous Infrared Galaxies (ULIRGs,~\citealt{Houck1985}) that have dramatically changed our understanding of the energy balance of the Universe. Through ALMA's fine angular resolution leveraging interferometry, the innermost regions of planet-forming disks where Earth-like planets may be found have been revealed~\citep{Rampinelli2024}. {\bf By opening up precision angular-resolution astronomy to the far-IR band, which cannot be effectively accessed from the ground, a far-IR interferometer provides the generational leap in observing capabilities that will similarly uncover new phenomena that re-write our understanding of the cosmos.} In these efforts, the broad and simultaneous spectral coverage of the double-Fourier technique is well suited, having substantial capabilities that are wholly unreliant upon priors.

As we aim to address the enduring questions of humanity, the successes of the Event Horizon Telescope (EHT) and Very Large Telescope Interferometer (VLTI), coupled with mission proposals such as Darwin, TPF-I, LIFE, and those listed here (i.e., SPICE/SPIRIT, SPECS, FIRI) are part of a cross-wavelength push to angular resolutions that exceed the capabilities of single-aperture telescopes. Indeed, interferometry from space-based platforms represents a potential and impending generational shift in astrophysical observatories. RoadMap2013~\citeyearpar{Roadmap2014} notes, ``Several of the notional missions listed in our roadmap rely on interferometry to answer key science questions, from radio to X-rays. All notional missions in the Visionary Era are interferometers, and technology maturation of interferometric techniques is highly relevant to realizing this science vision.'' {\bf Owing to the abundant discovery space of far-IR observations with a fine degree of angular resolution and the relative relaxation of constraints on optical systems that the longer wavelengths of the far-IR afford, we see far-IR interferometry as the most sensible first step in bringing about this generational shift.}  

\clearpage

\section{Mission concepts}
\label{sec:concepts}

The need for a facility capable of addressing the open questions in astronomy outlined above has led to far-IR interferometer mission concepts developed by international teams that have been proposed to both NASA and the ESA. In this section, we outline some of the most advanced mission concepts. Table~\ref{tab:concepts} provides a summary of the key mission capabilities and budget scale associated with each of the discussed concepts.

\begin{table}
    \centering
    \caption{A summary of general mission parameters for the concepts discussed in Section~\ref{sec:concepts}. The values are taken from the published works that are referenced in the text of this section.}
    \begin{tabular}{p{1.7cm}|p{3.6cm}|C{1.4cm}|C{1.8cm}|C{1.5cm}|C{2cm}|C{1.9cm}}
         Concept & System Architecture & FoV (arcmin) & Resolving Power ($R$) & Spectral Band ($\mu$m) & Maximum Baseline\,[m]/ Angular resolution [arcsec at 200\,$\mu$m] & Cost regime\\
         \hline \hline
         SPECS & Formation-flying with 2 collectors. Double-Fourier & 1 & 1000--3000 & 40-640 & 1000/0.02 & NASA Flagship \\
         ESPRIT & Formation-flying with 4 collectors. Heterodyne/Correlation & 0.1 & $>10^6$ & 50--600; instant. $\Delta\nu=4$\,GHz & 50/0.4 & ESA Large Observatory\\
         SPIRIT/ SPICE & Connected structure with 2 collectors. Double-Fourier & $>1$ & Up to $>3000$ & 25--400 & 30/0.7 & NASA Probe \\
         FIRI (straw-man) & Formation-flying with 3 collectors. Double-Fourier with single-dish Heterodyne & 1 & Up to 5000 and $10^6$ & 25--385 & 1000/0.02 & ESA Large Observatory\\
         FIRI (FISICA Study) & Connected structure with 2 collectors. Double-Fourier & $>1$ & Up to $>3000$ & 25--300 & 30/0.7 and 100/0.2 & ESA Large Observatory\\
    \end{tabular}
    \label{tab:concepts}
\end{table}

%\subsection{ESPRIT} Still deciding here
%Passively cooled since less sensitive (heterodyne). Long coherence length from narrow bandwidth. Might not include.

\subsection{SPECS}
The Submillimetre Probe of the Evolution of Cosmic Structure (SPECS; \citealt{Mather1998}) represents an early and highly ambitious mission concept for a far-IR interferometer. SPECS was added to the Roadmap for Astronomy and Physics in the 2000 Decadal Report from the USA National Academy of Sciences' Astronomy and Astrophysics Survey Committee~\citep{NASA2000Decadal}. This was met with the recommendation from the consensus of the participants at the ``Second Workshop on New Concepts for Far-Infrared/Submillimeter Space Astronomy'' held in 2002 that following an actively cooled large single aperture telescope, a ``second step'' should be pursued ``to achieve Hubble Space Telescope-class [spatial] resolution in the far-IR/sub-mm, which would require an imaging interferometer with a 1\,km maximum baseline''~\citep{Leisawitz2004}. In this configuration, SPECS was intended to take advantage of technologies developed by a large actively cooled single-aperture ``first step'' observatory, including: large format detector arrays and accompanying readout, JWST-like deployable multi-layer Sun shields, and cryogenic systems capable of actively cooling a large mirror. During its study, SPECS evolved from an array of 3 light collectors with 4-m\, mirrors arranged in a triangle about a central beam combiner employing formation flying to spatially sample the $u,v$ plane to later consider a tethered and/or truss structure with 1--3 simultaneous baselines~\citep{Harwit2006}. 

\begin{figure}
    \centering
    \includegraphics[width=0.5\linewidth]{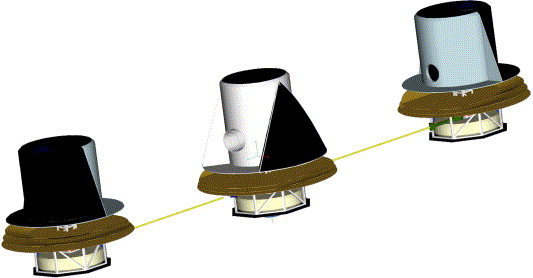}
    \caption{A rendering of the proposed SPECS architecture. Image Credit:~\citet{Harwit2006}.}
    \label{fig:placeholder}
\end{figure}

SPECS eventually settled on a design with two light-collecting telescopes and a central beam-combining instrument leveraging the double-Fourier technique. This technique takes full advantage of the sensitivity of a space platform through direct detectors over a quantum noise-limited heterodyne solution (though heterodyne spectroscopy was considered as an option to be integrated into one of the light collectors/telescopes). The double-Fourier technique was independently proposed by \citet{1988Mariotti} and \citet{Itoh86} as a method to provide high-resolution spatial and spectral astronomical observations over a broad spectral band. A double-Fourier interferometer performs direct interferometry by combining the light from two or more light-collecting telescopes into a central beam-combiner with a scanning delay line~\citep{Itoh86}. The spatial information of the observed field is encoded in an observation as light is collected from discrete positions/separations of the light collectors in space (sampling the $u,v$ plane). Through the delay line, interference signals are obtained in which the temporal coherence of the observed field is measured (as in Fourier Transform spectroscopy) along with its spatial coherence for different pairings of light collectors with some separation (or baseline) between them. From the interference measurements, the image of the field and its spectral information is recovered as a hyperspectral ``data cube'' through a three-dimensional Fourier transform. As a double-Fourier interferometer employs direct detectors, it has an inherent sensitivity advantage over heterodyne techniques, which are unavoidably limited by quantum noise. From its proposed design, SPECS became a NASA-commissioned ``vision mission'' study~\citep{Harwit2006,2007SPECS}.

\subsection{ESPRIT}
The proposed Exploratory Submm Space Radio-Interferometric Telescope (ESPRIT) concept provides an example of a different approach to far-IR interferometry that leverages heterodyne/correlation architectures (similar to ALMA)~\citep{2007Wild,2008Wild}. Initially, ESPRIT was envisioned as 6 (or more) light collectors with 3.5\,m mirror diameters in a free-flying formation. Such an arrangement was designed to achieve 15 simultaneous baselines of up to 1\,km maximum length with constant motion between the collectors to sample the $u,v$ plane. Owing to its heterodyne spectroscopy, ESPRIT would be capable of both high spatial and spectral resolution measurements. Interferometry through heterodyne spectroscopy can exploit large coherence lengths (owing to its small instantaneous spectral bandwidth), separate recording and subsequent correlation of the signals received by each of its light collectors, and a relaxation on the relative positional control requirement of the facility since, in practice, knowledge of only the path length between these is sufficient~\citep{Leisawitz2013}.

As a consequence of the limited sensitivity of heterodyne receivers due to the quantum noise limit, ESPRIT was not intended to employ actively cooled apertures~\citep{2007Wild}. Indeed, though the detector and mixer backend would require sub-4\,K temperatures, the light collectors themselves would only need a temperature of $\sim$70\,K to reach the sensitivity limit of the heterodyne receivers. Owing to the spectral resolving powers above $10^6$, limited instantaneous bandwidth ($\sim$4--16\,GHz), and limited sensitivity of heterodyne-based architectures, there is an appreciable disparity between many of the scientific aims, the discovery space, and the reliance upon priors of a heterodyne-based interferometer and those of a direct-detection interferometer (such as the double-Fourier technique). While a full discussion of these are beyond the scope of this paper, ESPRIT serves as a useful point of comparison. Here, we have highlighted a few of the technical differences between each technique. More substantial discussions and studies of these trade-offs can be found in \citet{FISICATradeoffs, FISICAFinal, Sauvage2013} and~\citet{2007Wild}.

ESPRIT was further developed as a concept for the ESA Cosmic Vision (2015--2025), eventually settling on a more modest design with 4 light collectors and baselines of up to 50\,m~\citep{Sauvage2013}. Space-based heterodyne interferometer concepts in the sub-mm and far-IR continue to be developed through the Space-based High-resolution Array for Radio astronomy and Physics (SHARP) concept and the Space Array initiative.

\subsection{SPIRIT/SPICE}
\begin{table}[b]
    \centering
    \caption{Selected performance requirements for subsystems to stabilise the SPIRIT optics bench and ensure phase stability of the interferometer. Many of these are design specific and interrelated. These are shown only to signpost engineering requirements for a modest far-IR interferometer. See~\citet{Wilson2007}.}
    \begin{tabular}{l|l}
         Component & Precision \\
         \hline \hline
         Baseline length precision & 10\,mm over 40\,m \\
         Baseline direction precision & 30\,arcsec \\
         Bearing precision (between collector and telescope) & 10\,arcsec \\
         Secondary mirror pointing & 0.1\,arcsec over $\pm$ 3\,deg\\
         Mirror steering from light collector to beam combiner & 1\,arcsec over $\pm$2\,deg \\
         Mirror steering within combiner & 0.3\,arcsec over $\pm$ 1\, deg \\
         Delay line zero path difference centring mechanism & 50\,$\mu$m over $\pm$ 3\,mm at 2\,Hz\\
         Path length compensator & 0.25\,$\mu$m over $\pm$ 2.5\,mm \\
         Telescope pointing mechanism & 10\,arcsec over $\pm$ 3\,deg
    \end{tabular}
    \label{tab:spiritspecs}
\end{table}
Coupled with the SPECS proposal, a near-term interferometry pathfinder mission was proposed. The Space Infrared Interferometric Telescope (SPIRIT) was proposed as a structurally connected double Fourier interferometer consisting of 2 light collectors, both $\sim$1\,m in diameter, and a central beam combiner~\citep{SPIRIT2007,SPIRITPropBook}. The two actively cooled light collectors travel symmetrically with a maximum baseline of 36\,m while the telescope rotates to sample the $u,v$ plane. The observing time was intended for the telescope rotational period to be on the order of $\sim$40\,minutes, the extension/retraction of the baseline to be on the order of hours. In total, measurements at each $u,v$ plane position were anticipated to be on the order of minutes, building to full imaging spectroscopic observations across the full 1\,arcminute field of view in up to $\sim$24 hours, largely depending on the spatial resolution needs for the science case. The maximum optical delay provided in the beam combiner was set to be $\pm$1\,m to allow for both $R\sim3000$ and geometric delay across the field of view.

\begin{figure}
    \centering
    \includegraphics[width=0.5\linewidth]{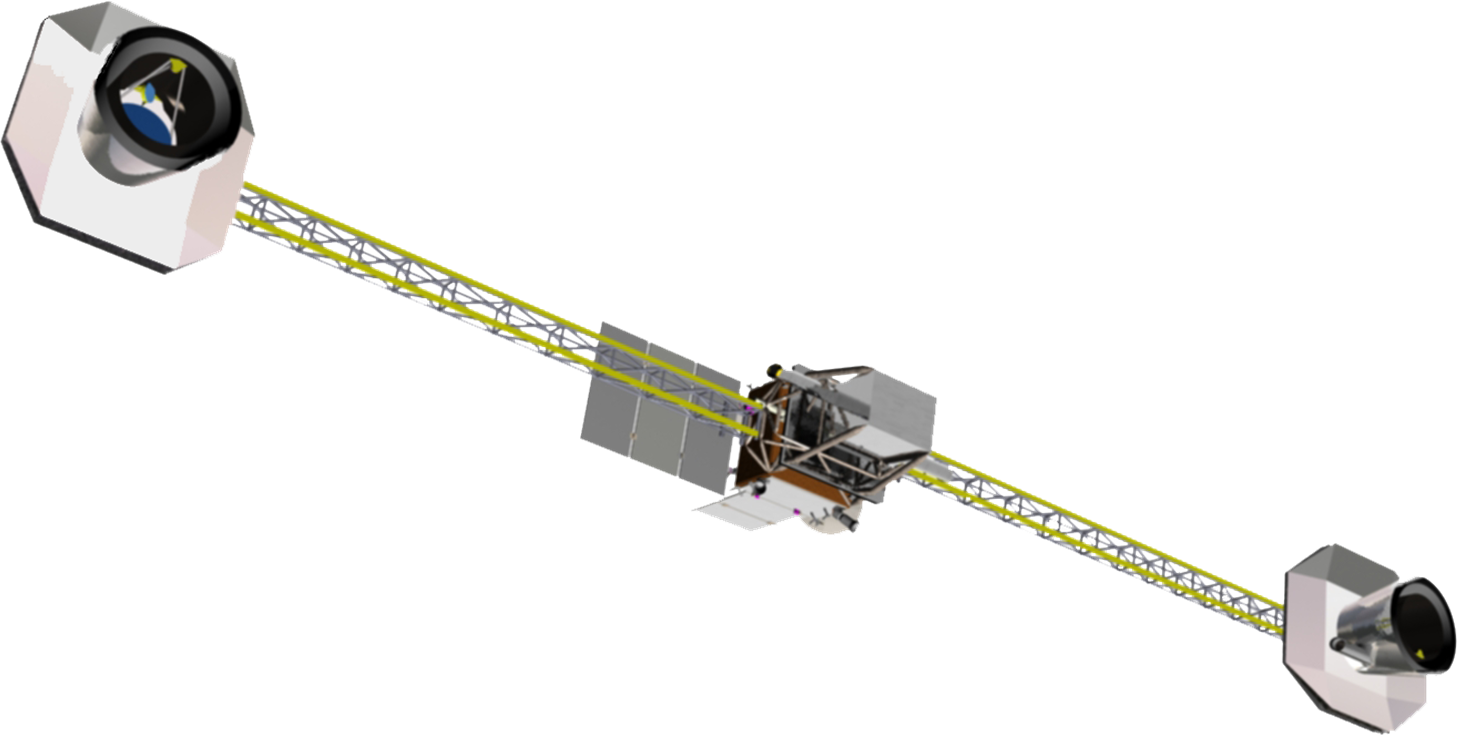}
    \caption{A rendering of the SPIRIT/SPICE mission concept. Image courtesy of the SPICE instrument and simulation team.}
    \label{fig:SPICE}
\end{figure}

SPIRIT was recommended in the community plan that emerged from the ``Second Workshop on New Concepts for Far-Infrared/Submillimeter Space Astronomy''~\citep{Workshop2004} as the logical first step towards a SPECS-like mission. Studied as a NASA ``Origins Probe'', SPIRIT is perhaps the most mature concept for a far-IR interferometry mission. In the Origins Probe study, significant progress was made in determining the subsystem capabilities, tolerances, and internal metrology to maintain good fringe visibility and phase stability in the interferometer~\citep{Hyde2007}. In general, stability on the order of $\lambda/10$ is needed~\citep{FISICAInterf}; in Table~\ref{tab:spiritspecs}, we note the specific mechanical tolerances that were required for the SPIRIT concept to meet this target. Through beam steering mirrors, changes to the angular direction of the beam from each telescope are compensated, and the optics bench is stabilised. Meanwhile, additional scanning mirror mechanisms within the delay line are used to compensate for unintended changes in the separation between light collectors. A separate ``metrology beam'' at shorter wavelength is used within the delay line to decouple changes in optical path difference from the light collectors from that of the delay line. Furthermore, SPIRIT also coupled in measurement of near-IR bright guide stars with a separate interferometer using the same primary optics to serve as a phase reference in the optical field, providing precise measurements of any path difference between each arm of the interferometer. Coupled with near-IR phase referencing, SPIRIT intended to make use of a similar laser metrology scheme as that developed for the Space Interferometry Mission (SIM, \citealt{Peterson1996SIM,Allen1997,Halverson2000}) to track the relative positions of the two light-collectors, adapted to the much relaxed requirements for an interferometer operating in the far-IR. This metrology scheme was found to be within a relatively modest technology development program and saw significant development through ballon-based pathfinder experiments (see \ref{subsec:Balloon} below).

Since its initial proposal, the SPIRIT design has been further developed and studied through efforts to build towards a potential proposal for the NASA Astrophysics Probe Explorer Program (2023). The Space Interferometer for Cosmic Evolution (SPICE) ultimately was not put forward but has been proposed recently in NASA's 2026 Ad ASTRA incubation program as a base platform that could be paired with complementary science instruments (such as a single-dish heterodyne receiver, or a mid-IR nulling interferometer pathfinder) for a potential ``flaglet''-class mission (between 1-2 billion USD). A detailed Science Traceability Matrix was developed by the SPICE team in 2022.

\subsection{FIRI}
A Far-IR Interferometer (FIRI) was proposed to ESA in the 2009 Cosmic Vision (2015--2025) call following its identification as priority in the ESALB 2005 symposium~\citep{FIRI2009}. Initially proposed as an ambitious kilometre-long baseline, FIRI sought to address all of the science goals outlined in table~\ref{tab:sciencecases} through the combination of both a direct-detection interferometer and a single-dish heterodyne system incorporated across three light collectors with a central beam combiner. In a similar fashion to SPECS, each light collector consisted of a 3.5\,m actively cooled afocal off-axis telescope and was to be connected via tethers or formation flying. A smaller structurally-connected concept, similar to SPIRIT, was proposed alongside the larger FIRI proposal and was selected by ESA for further study through a Technology Reference Study~\citep{FIRICDF}. This design employed two 1\,m collectors separated from a common hub, containing the beam combiner, by means of a mechanical structure that permits the separation/baseline between the collectors to be varied. As with SPIRIT, spatial sampling of the $u,v$ plane would be achieved through the adjustable baseline and rotation of the interferometer.

\subsection{Conclusion}
Through the mission concepts and studies outlined above, the double-Fourier interferometer (DFI) has emerged as the most viable pathway to sub-arcsecond spatial resolution in the far-IR. Indeed, the imaging advantages of the DFI, with its broad spectral band coupled with medium-to-high resolution capabilities are well-matched to the wide discovery space that a generational leap in spatial resolution would provide. Furthermore, many key DFI subsystems have a high technology readiness, most notably cryocooler technology and exquisitely sensitive detector arrays. Furthermore, many of these technology areas are synergistic with PRIMA and stand to continue towards TRL9, as PRIMA develops towards flight.

\section{Pathfinders and Demonstrations}
\label{sec:pathfinders}

\subsection{Testbeds}

\subsubsection{The Wide-field Imaging Interferometry Testbed}
The Wide-field Imaging Interferometry Testbed (WIIT) was developed at the NASA Goddard Space Flight Center in an effort to validate and experiment with the wide-field imaging capabilities enabled by the double-Fourier technique and build toward the proposed SPECS and SPIRIT missions~\citep{Leisawitz2003, Leisawitz2012}. Indeed, WIIT was built as a 1:150 scale model of SPIRIT operating at optical wavelengths to take advantage of readily available detector and opto-mechanical components. 

With WIIT, a series of test scenes were collimated into a large beam incident on the testbed interferometer. Via two moving mirrors, the wavefront from the test scene was sampled, and $u,v$ plane coverage was obtained by rotating the test scene rather than the interferometer and adjusting the separation of these two input mirrors perpendicular to the incoming beam.

WIIT represents a significant step in demonstrating not only the double-Fourier technique, but in particular its capabilities to deliver hyperspectral images of simulated astrophysically interesting sources with a wide field of view~\citep{Leisawitz2016,Roser2016}. These ranged from simple broadband binary point sources with and without a spectral emission line to a simulated far-IR wide-field extragalactic observation with several galaxies with different morphologies and redshifts as well as contaminant foregrounds (Galactic cirrus and zodiacal light). Furthermore, WIIT provided a robust demonstration of the metrology and control systems necessary for a double-Fourier interferometer to compensate for the relative alignment of both input apertures/mirrors and to precisely monitor both the position of these and the scanning delay line mechanism~\citep{Douglas2003, Rinehart2007}.

\subsubsection{The Multi-Fourier Transform Interferometer}
The Multi-Fourier Transform Interferometer (MuFT) developed by \citet{2006Ohta} incorporated polarisation into double-Fourier interferometry, similar to a Martin-Puplett interferometer~\citep{1974MP}, allowing measurement of spatial and spectral information alongside four Stokes parameters. Furthermore, the Martin-Puplett interferometer enables the subtraction of common fluctuations using orthogonal polarising detectors, so that background and temperature fluctuations can be suppressed, as demonstrated in \citet{Luo2008}.

The MuFT, having a particular emphasis on the sensitivity advantage of direct detection interferometry over heterodyne techniques, operated in the sub-mm to far-IR spectral band with a single composite bolometer while employing a similar optical design to WIIT (only with the addition of wire-grid polarisers). The MuFT measured the two-dimensional complex visibility of different aperture masks with differing hole patterns~\citep{2007Ohta}. The MuFT was further developed towards astrophysical observations on the sky presented by~\citet{2007Ohta2} and \citet{Luo2008}. Furthermore, it demonstrated a technique for large-dynamic-range imaging for sources for which their spatial distribution is decoupled from spectral frequency. Through this technique, a large $u,v$ plane coverage is obtained from a small number of baseline combinations through co-adding the images obtained at different frequencies~\citep{2006Ohta}.

\subsubsection{The Far-Infrared Space Interferometer Critical Assessment}
Coupled with ESA's study of FIRI, the European Commission supported the Far-Infrared Space Interferometer Critical Assessment (FISICA) as part of its seventh framework program for research development and demonstration activities~\citep{FISICAScience}. The consortium, consisting of researchers across the UK, EU, and Canada, conducted a dedicated study effort to assess the technology readiness level (TRL) of key subsystems and components for far-IR interferometry and to engage in further TRL development from 2012 to 2015. 

From the FISICA study, two testbeds were developed to study the double-Fourier technique. The first, discussed in \citet{2012Grainger}, operated in the sub-mm to terahertz regime and employed a single silicon bolometer. As in the previously discussed testbeds, light from a large collimated beam was sampled by two apertures. However, the optical design employed only a single moving mirror to achieve baseline separations from a fixed mirror. Using a single polariser, the testbed could be easily reconfigured to behave as a single-port Fourier transform spectrometer, allowing for an easily accessible zero-baseline mode. This testbed was successful at measuring simple test scenes consisting of one or two slits. The second testbed was designed to operate at shorter mid-IR wavelengths to take advantage of off-the-shelf imaging focal planes to test off-axis effects but did not progress towards spatial interferometric measurements prior to the end of the program~\citep{FISICAFinal}. Coupled with the testbed instruments, the FISICA study included notable efforts to test the performance of cryogenic delay lines and associated metrology~\citep{FISICADelay}. 

In complement to the technology development, the FISICA study made significant progress in the development of software tools and algorithms to simulate far-IR interferometers and process their data. The Python Far Infrared Instrument Simulator (PyFIInS) was developed to provide an end-to-end simulation of FIRI, which continues to be developed~\citep{FIINS, 2025Spencer}. Additionally, significant progress was made in community development through three international community far-IR interferometry workshops.

\subsubsection{University of Lethbridge Double-Fourier Testbed}

Building on the heritage of the aforementioned testbeds, the University of Lethbridge (Canada) has developed a double-Fourier interferometry testbed operating in the 300--2100\,GHz spectral regime, capable of wide-field imaging spectroscopy, and employing beam-dividers similar to those envisioned for a SPICE-like mission~\citep{Carole2007}. Baseline aperture separation is achieved with a stationary mirror and a single moving mirror. The delay line follows a Mach-Zehnder-style FTS, as with the \textit{Herschel} Spectral and Photometric Imaging Receiver (SPIRE) FTS~\citep{Griffin2010SPIRE}. The testbed is coupled to a custom array of 25 feedback-controlled superconducting transition edge sensors designed to demonstrate the wide-field imaging of a double-Fourier interferometer in the sub-mm/far-IR~\citep{DFIJAtis2025}. As of writing, this testbed has been successful in providing spatial and spectral interferometric measurements of spatially simple test scenes with some spectral variation using only a subset of the detector array~\citep{JeremyTestbed,JeremyTestbed2}. Measurements exploiting the full detector array and analysing more spatially complex test scenes are anticipated in the near future.

\subsection{Balloon Pathfinders}
\label{subsec:Balloon}

\subsubsection{The Far-IR Interferometric Telescope Experiment}
The Far-IR Interferometric Telescope Experiment (FITE) was conceived as a balloon-borne Michelson stellar interferometer capable of resolving the angular size of spatially simple objects to within a 4 arcsecond spatial resolution at 150\,$\mu$m, a band that is inaccessible to ground-based observations~\citep{2010Shibai}. Employing a fixed 8\,m baseline, FITE aimed to use a linear array of detectors to measure the spatial interference pattern from targets with a spatial distribution similar to a uniformly illuminated disc~\citep{2010Kato}. This interference pattern could then be compared against measurements of point sources to determine the angular size of the disc-like object. 

Misalignment between both light collectors of the interferometer due to mechanical deformations arising from thermal contraction and deflections caused by the interferometer's motion during flight could be compensated through beam-steering mirrors and a wavefront sensor~\citep{2012Sasaki}. In laboratory tests, the control systems for FITE demonstrated within 11.6\,arcsecond parallelism of the two beams from the light collectors and within 80\,$\mu$m agreement between the foci~\citep{2012Sasaki}. Despite this success, FITE did not develop beyond technology demonstration. 

\subsubsection{The Balloon Experimental Twin Telescope for Infrared Interferometry}
Another notable far-IR interferometry pathfinder was the Balloon Experimental Twin Telescope for Infrared Interferometry (BETTII), which was selected for funding through the NASA ROSES/APRA program in 2010 \citep{2010Rhinehart,2012Rhinehart,Rinehart2014}. Also employing a fixed 8\,m baseline between two light collectors, BETTII was designed to achieve sub-arcsecond spatial resolution in the 30-90\,$\mu$m band leveraging the double-Fourier technique. Thus, BETTII was poised to provide imaging spectroscopy of star formation regions and AGN with an otherwise unprecedented spatial resolution in the far-IR band that is inaccessible from the ground while advancing the system-level TRL of far-IR DFI.

BETTII employed a similar control and metrology scheme to the concept that was devised for the SPIRIT mission concept~\citep{2014Rizzo}. Here, we provide an overview of these systems. Using a near-IR camera fixed on a guide star and a series of inertia wheels, a control loop was designed to fix the pointing of the payload to within 15\,arcseconds of the guide star. Fine control of the pointing is achieved through fast steering tip/tilt mirrors that are placed at the pupil of each arm of the interferometer and a near-IR detector that maintains the location of the two images of the guide star (one from each arm) such that overlap in the science channels is maximised. A secondary delay, placed in only one arm of the interferometer and secondary to the delay line used to probe the interference pattern in the double-Fourier arrangement, is used to lengthen or shorten one arm to compensate for any phase errors from attitude control. To correct for any drifts in phase, fringes from a point source must be frequently referenced. To provide this ``fringe-tracking,'' BETTI separated out near-IR light from each arm of the interferometer and measured the central interference fringe arising from these near-IR beams to achieve a high sensitivity to small differences in the path length from each arm of the interferometer. Fine control of relative path lengths of each interferometer arm was then compensated through another moving mirror within the cold optics of the interferometer. Positional metrology of the delay line within the DFI was obtained via a laser differential interferometer. A similar system was also used to measure path length variations between the central beam combiner and the two light collectors~\citep{2010Rhinehart}. Additionally, BETTI underwent notable efforts to model and compensate for mechanical vibrations and deflections in the interferometer's boom structure~\citep{2012Rhinehart,Rhinehart2018,Rinehart2014}.

Through laboratory testing, the pointing control loop accuracy necessary for operations was demonstrated for BETTII~\citep{Rinehart2014}. In 2017, BETTII underwent a successful engineering flight which demonstrated  maintenance of optical alignment through launch, more than sufficient closed-loop control from coarse pointing systems, and validation of the near-IR detector array~\citep{Rhinehart2018}. BETTII also demonstrated the general thermal, electrical, and mechanical performance necessary for success. During its flight, BETTII unfortunately experienced a fault that separated it from its parachute during descent, resulting in near-total loss of the payload.

\section{Enabling technologies and their readiness}
% SPICE FAQ pages. Time of Observation for targets, etc.
% Engineering and CONOPS. 
%TODO: Is there room to discuss civil application? optical communication?
\subsection{Deployable boom technology}
Many of the mission concepts outlined in Section~\ref{sec:concepts}, particularly those which are most promising for near-term implementation, are reliant on a rigid connecting structure that is deployed from some central launch bus. The stability of this connecting structure, particularly with respect to transverse vibrational modes in which the light collectors would be displaced in opposing directions, must be such that phase stability can be maintained through metrology and compensating control systems. This requirement is, of course, a function of the maximum baseline achievable by the interferometer.

The simplest option is a structure with hinges, enabling it to fold into a configuration compatible with the dimensions of a rocket fairing. The light-collecting telescopes of an interferometer can move along rails attached to the structure. Commercial launch vehicles available today or in the near future could accommodate hinged, structurally connected interferometers with baseline lengths approaching 100\,m. However, space-qualified deployable boom technology progressed in parallel with launch vehicle developments during the past several decades.

The Nuclear Spectroscopic Telescope Array (NuSTAR) mission represents significant heritage for deployable booms for precision optical assemblies in space~\citep{Harrison2013}. NuSTAR achieved an extended separation of 10\,m between the optics at either end of the boom and lateral deflections of only a few millimetres. This was coupled with laser metrology capable of $\sim$10\,$\mu$m positional tracking resolution. A similar deployable boom was used on the Shuttle Radar Topography Mission (SRTM)~\citep{Farr2007}. The NuSTAR and SRTM booms were stowed in compact configurations for launch and unfurled in space. An interferometer’s telescopes could be fixed at the ends of booms deployed from canisters, perhaps obviating the need for a rail for telescope transport. While booms designed for single deployment or a small number of retractions after initial deployment have attained the highest technology readiness level (TRL 9), a long boom (up to 50 m) that can survive, say, thousands of deployments and retractions has not yet been demonstrated. This approach, as opposed to a hinged structure, warrants further examination for space-based interferometry.

It should be noted that time between now and two of the most substantive studies into the feasibility of a far-IR space-interferometer, FISICA and SPIRIT, has seen the successful deployment of the James Webb Space Telescope (JWST). This has dramatically changed our perception of what is possible in terms of deployable structures and vibrational suppression from a space telescope~\citep{Reynolds2004, Menzel2023}. Thus, there is substantial opportunity for future far-IR interferometer concepts to study how best to implement JWST's advances in these areas. %Indeed, many of the rigid deployment structures on-board JWST, of order 7-10\,m in length, 

\subsection{Wavefront performance and fringe tracking}
Though interferometry allows a significant easement in the size of primary mirror for a facility to achieve a fine degree of angular resolution, the mirror sizes that have been discussed in Section~\ref{sec:concepts} are of a size that warrants consideration for a space-based observatory. To minimise fringe visibility loss, the SPICE mission targeted a wavefront loss of under $\lambda$/60. The \textit{Herschel} Space Observatory's primary mirror, being 3.5\,m in diameter, at the time of writing represents the largest monolithic mirror on a space-based observatory. Without any active control of the mirror shape, it provided a total wavefront error of less than 6\,$\mu$m~\citep{Pilbratt2010}, which would provide sufficient performance at 350\,$\mu$m for an interferometer mission. Further progress is expected; as PRIMA progresses towards flight, it will provide a steady development program toward $\sim$2\,m mirrors with wavefront error performance under 500\,nm~\citep{Glenn2025PRIMA}. In spite of this promising technology development foundation, an interferometry mission must account for the performance of multiple light collectors. Indeed, either comparable deformations on all of the cooled apertures or some way of assessing and/or compensating for these is necessary and the large- to mid- scale wavefront error distributions must be considered. As such, the feasibility of on-board wavefront sensing and/or adaptive optics systems is worthy of study.

The use of a shorter wavelength interferometer to provide a phase reference, as was proposed for both SPIRIT and FIRI, does introduce an additional concern that the light collector mirror quality is sufficient for some degree of near-IR interferometry. A detailed study of error budgets in the SPIRIT design in~\citet{Hyde2007} determined that a visibility of 25\% is all that is necessary to achieve fringe tracking via a near-IR interferometer and that this could be reasonably achieved within the budget of an Origins Probe-class mission.

\subsection{Spacecraft structural metrology}
Fringe tracking via a shorter-wavelength interferometer, as has been discussed, does raise some concerns as to whether the instantaneous feedback necessary to compensate for misalignments and structural bending is possible. Also, the need for a bright guide source does impose some limitation on the fields of observation available to the interferometer. Whether in complement to, or in partial replacement of short-wavelength fringe tracking altogether, substantial metrology of the relative positions and pointings of the interferometer's light collectors is necessary. As such, the metrology of spacecraft structures has received significant attention.

Laser metrology systems suitable for interferometric application in space have seen significant developments over the past decades. In development efforts for the SIM, a laser metrology system capable of less than 10\,$\mu$m accuracy for absolute positional measurements and less than 100\,pm accuracy for relative positional measurements over lengths of order $2$\,m was demonstrated via laboratory testing~\citep{Dekens2004}. This far exceeds the requirements for a modest ($\sim30$\,m) interferometer noted in Table~\ref{tab:spiritspecs}, albeit at a smaller scale. Recently, the PROBA-3 mission has been successful in maintaining $\sim$150\,m formation flying with mm-level positioning through a combination of propulsion thrusters, reaction wheels, and its laser metrology system~\citep{Loreggia2020,Zhukov2025,TarabiniCastellani2013}. This mission, like SIM, provides metrology with the tighter precision requirements for shorter wavelengths. As the TRL for laser-based metrology systems for space-based interferometers is relatively high, is likely that only modest development will be needed to develop a similar system tailored for a far-IR DFI. Furthermore, alternate approaches that leverage accelerometers have also been studied as part of the FISICA program~\citep{Iafolla2014}.

\subsection{Cryogenic metrology}
Micron-level metrology precision is required for far-IR interferometry. As noted above, through laser metrology, precise measurements of the relative position of optical components over distances of order $\sim$10\,m have been demonstrated. Equally important is precise metrology of the delay line and path compensation mechanisms within the beam combiner's cryogenic optics bench. Substantial developments have been made in preparation for future far-IR missions. Notably, \citet{2026Adam} have developed a laser metrology system capable of simultaneously measuring the displacement of up to 8 cryogenic delay lines with $\sim$50\,nm accuracy~\citep{2026Matt}. Such a system is more than sufficient to provide internal metrology for the beam combiner of a far-IR DFI. %Furthermore, its multi-axis nature may see further development towards baseline scales of order $\sim$30\,m.

\subsection{Cooled apertures and focal planes}
A significant capability a space-based DFI enables is the improved sensitivity that can be obtained through direct detectors over heterodyne architectures. To fully exploit this sensitivity across the full band of the far-IR angular resolution gap, apertures of the light collectors must be cooled to temperatures of $\sim$4\,K, with sub-Kelvin cooling needed for the detectors. While such a task is daunting for a single aperture facility aiming for a similar level of angular-resolution through a large primary mirror, the cooling requirements of a modest SPIRIT/SPICE-like instrument with metre-class primary mirrors are in-line with what has and continues to be developed for the SPICA and PRIMA missions~\citep{Sugita2010,2016Shinozaki,Chen2026} and a natural extension of the cryocooler technology developed for JWST’s MIRI instrument~\citep{Wright2023}. Several options are available from vendors in the US and Japan, as described in the Origins Space Telescope Technology Development Plan~\citep{OriginsStudyTeam2019}. As such, the necessary 4\,K cryocooler technologies are already of a high TRL, with steady progression up to 9 anticipated through PRIMA's development program. 

As in a single aperture facility, the sensitivity of an interferometer scales with light collecting area. In interferometers, this is often achieved by increasing the number of separate light collectors/apertures as this simultaneously increases sensitivity and instantaneous $u,v$ plane coverage. For a background-limited far-IR interferometer, increasing the number of light collectors naturally requires increasing the number of $\sim$4\,K cryocoolers (and their redundant systems required for risk mitigation in a space-based mission). This represents an appreciable increase in power draw, launch mass, and the mass of the light collectors, which in many cases must be moved throughout an observation to set the baselines of the interferometer. This is, in large part, the reason the majority of DFI concepts discussed throughout this paper have converged to designs with only two light collectors. Progress in reducing both the mass and power requirements of $\sim$4\,K cryocoolers has the potential to dramatically improve on observation acquisition times of space-based far-IR DFIs by enabling more light collectors. 

Cooling far-IR detector arrays to the necessary sub-K levels ($\sim$50\,mK) can be achieved via a continuous adiabatic demagnetization refrigerator (CADR, \citealt{Shirron2002,Tuttle2017}). This technology also has a clear pathway to TRL 9 through PRIMA. Again, the Origins Space Telescope had similar needs, and potential solutions are described in the Origins Space Telescope Technology Development Plan~\citep{OriginsStudyTeam2019}.

\subsection{Post-dispersion}
The spectral information in a DFI is a consequence of the scanning delay line, which operates as a normal FTS in large part. FTSs hold significant advantages with respect to both optical throughput and multiplexed imaging spectroscopy when compared to traditional dispersing spectrometers. In an FTS, all spectral components are measured at all times by each detector, i.e., each imaging detector in an array simultaneously measures a full spectrum. As a consequence, the large spectral band incident on each of its detectors results in a significant increase in photon noise. With modern far-IR detectors capable of NEPs below $10^{-19}$\,W\,Hz$^{-0.5}$, the multiplexing advantage of the FTS comes with a significant penalty to the instrument's sensitivity. This penalty can be mitigated without limiting the overall bandwidth by adding a low-resolution dispersive spectrometer at the output of a higher-resolution FTS. The low-resolution post-dispersing spectrometer serves to break up the bands at the focal plane, reducing noise and improving the instrument's sensitivity~\citep{Naylor2022}. An FTS coupled to a grating post-dispersing spectrometer has seen significant interest and was adopted for the SAFARI instrument for the SPICA mission concept and is being developed for the FIRESS instrument on PRIMA~\citep{Naylor2022,Bradford2025}. DFI instruments are well-positioned to benefit from this technique.

It should be noted that the use of a grating spectrometer, while well-matched to point-like extragalactic single targets, would compromise the fundamental imaging advantages of the FTS within a DFI in many respects. In cases where slits are employed with the grating module, the natural throughput advantage of the FTS (containing only circular apertures) is compromised. Furthermore, in most practical cases, a substantial portion of the focal plane's spatial extent must be dedicated to spectral detection rather than imaging, thus sacrificing the simultaneous acquisition of spatial imaging and spectral information of the FTS (which then must be recovered through additional pointings). On-chip filterbank spectrometers are a burgeoning technology to provide compact spectral dispersion on the focal plane that can be packed into 2-D imaging focal planes that in many ways would provide an ideal post-dispersion device for an FTS~\citep{Me2026}. However, these are limited to frequencies below the superconducting bandgap of the materials used in the feedlines and filters from which these devices are constructed. As such, their implementation has currently been limited to mm to sub-mm frequencies ($\sim$100--500\,GHz)~\citep{Endo2012,Barry2022}. In principle, these devices are capable of reaching spectral frequencies of up to $\sim$1\,THz through materials such as NbN and NbTiN~\citep{Barry2014PHD}. %Filterbank spectrometers constructed in vacuum waveguides etched into silicon via deep reactant ion etching also provide a promising path to low-resolution spectrometer focal planes that could be used for post-dispersion ~\cite{SWISSS}. However, such devices would also be limited to low terahertz frequencies due to the challenges inherent in fabricating high-frequency waveguides.

\subsection{Detector technology}
A far-IR interferometer is well positioned to take advantage of current advances in both detector sensitivity and multiplexing. Through development programs for SPICA and PRIMA, both transition edge sensor bolometers (TESs) and kinetic inductance detectors (KIDs) have emerged as suitable direct detector technologies for background-limited far-IR astronomy from a space-based platform~\citep{2016Suzuki,2025Hailey}. While both technologies are suitable, KIDs are capable of achieving faster detector time constants than TESs by a significant margin~\citep{2018Mauskopf,Griffin2016}. This is particularly advantageous for double-Fourier interferometry as spectral information and $u,v$-coverage require adequate sampling rates while both the delay line and spacecraft are in continuous motion~\citep{Benford2007}.

Far-IR interferometer mission concepts that have been studied and proposed (see Section~\ref{sec:concepts}) do not require performance beyond the sensitivity and multiplexing density requirements of both PRIMA and SPICA, and thus the detectors necessary for a successful far-IR interferometry mission are already of a very high TRL (with a clear path towards flight heritage with the anticipated success of PRIMA). However, interferometric measurements do introduce particular technical challenges for current far-IR detectors that must be fully addressed for the development of a far-IR interferometer. Notably, observations with dense $u,v$ coverage may necessitate longer time scales between calibrator measurements, which may exceed the time scale of correlated 1/$f$-noise signals.  Additionally, a rigid-bodied interferometer, such as has been proposed in the SPIRIT/SPICE mission concept, represents a large mechanical structure when compared to single aperture facilities. The vibrational environment and its effect on sensitive cryogenic detectors must be examined. Mechanical studies as part of the SPIRIT program have indicated that the large structure may be less prone to microphonics than that of a smaller single-aperture facility~\citep{Hyde2007,Budinoff2007}. 

%There has been some success in removing 1/$f$-noise from double-Fourier measurements by subtracting the time-ordered data from detectors placed at each of the two output ports from the interferometer (these being 180 degrees out of phase at each output of a well-balanced interferometer)~\citep{2007Ohta2}.

%System level demonstration of interferometry in SPACE. Hiroshi's japanese dmeonstration. What remains? Careful not venture too far into engineering. Note SIM metrology on KITE. Note relaxed requirements for far-IR.

%near IR optics. Interferomerty requirmeents relaxed. Just need adequate fringe visibility. Studied in SPIRIT. 

\section{Engineering notes, systems-level demonstrations, and potential pathfinders}
Just as with ground-based interferometers at radio, submillimetre, and optical/near-IR bands, interferometry from a space-based platform in the far-IR represents a significant shift in the observing strategy, operations, and engineering requirements from those in a single-aperture facility. Many of these are specific to particular designs and even to particular science targets. Some general investigations into the operational concepts for a far-IR interferometer were studied as part of the FISICA program~\citep{FISICACONOPS}. Further study will need to play an integral role in future mission-specific studies. The SPIRIT/SPICE mission concept provides such an example, being a relatively mature concept. The rotation of the connected structure was projected to complete a full circle of the $u,v$ plane at a given baseline in half an hour, with a field of regard being available for up to 60\,days~\citep{SPIRITPropBook}. The time to complete an observation would then scale based upon the density of baselines required to achieve a $u,v$ plane sampling necessary for the science goals of any given observation, coupled with the integration time needed. With this arrangement, SPIRIT was projected to have a mission lifetime of 5 years~\citep{Hyde2007,SPIRITPropBook}.

With ambitious new concepts for space-based facilities, smaller scale pathfinders and ground based testing are crucial. As we have detailed in Section~\ref{sec:pathfinders}, there has been significant progress in developing DFI techniques, key subsystems, and observation processing algorithms through several TRL-raising ground-based testbeds and balloon-based pathfinders. However, a full control systems demonstration with the necessary performance for an effective far-IR DFI mission has yet to be demonstrated in the far-IR band. Much of this can be advanced through ground-based testing. In particular, demonstrating the metrology and optical path compensation required to maintain phase stability of order $\lambda/10$ between each of the light collectors in a DFI (at large baseline separation) is a compelling opportunity~\citep{FISICAInterf}. Spatial and spectral information in the DFI are both measured via the scanning delay line, thus the phase stability control system must be coupled in with the delay-line metrology such that all spatial information can be recovered without substantially compromising the resolution requirements to achieve the science of a DFI mission. A testbed instrument, designed to provide this full systems-level demonstration of a stable DFI optical bench, has been proposed to the NASA Research Opportunities in Space and Earth Science 2026 program (PIs: S.~Ertel and J.P.~Scott).

It should be noted that there are also significant development efforts for direct-detection space-based interferometry pathfinders in the mid-IR for the purpose of nulling interferometry. These share several architectural similarities with a DFI mission (See \citealt{Quanz2026}). One notable example is the Large Interferometer For Exoplanets (LIFE) mission proposal, which is currently building towards a small-scale pathfinder mission proposal ($\sim$1.5\,m maximum baseline)~\citep{Glauser2026,Quanz2022}. Recently, the potential for a modest far-IR interferometer to also couple in a large-scale pathfinder for mid-IR nulling interferometry ($\sim$30\,m baseline) has been noted in a proposal to NASA's 2026 Astrophysics Strategic Technology and Research Accelerator (ASTRA) program (PI: L.~Mundy).

Owing to the technical challenges that must be overcome for sensitive observations in the far-IR (see Section~\ref{sec:science}), smaller scope (and cost) pathfinder missions for a far-IR DFI are a difficult prospect. Indeed, the requirement for powerful cryocoolers often precludes CubeSat and SmallSat platforms. DFI has not yet had a modern full-systems level demonstration from a space-based platform at any wavelength, thus there is opportunity for a thermal/mid-IR pathfinder to provide further TRL development for a far-IR space-based DFI. Such a pathfinder, leveraging these small platforms, was recently proposed by the Space-based Interferometer Feasibility (SPIFF) Project as part of the UK Space Agency (UKSA) Frontiers 2035 program~\citep{Ricketti2026}. Additionally, there is significant potential to leverage these small cost-effective platforms with Schottky-diode mixers to demonstrate space-based far-IR interferometry in general, these being sensitive enough to observe solar planets at terahertz frequencies without the need for substantial cooling efforts~\citep{Mehdi2017}. Two CubeSats could be connected or formation-flying with separate data acquisition and recording system. Such an experiment would obtain the first images of solar planets with a fine angular resolution at terahertz frequencies.

\section{Conclusion}
There is a wide array of science that can be addressed through an orders-of-magnitude improvement in the angular resolution of far-IR observations. In this paper, we have outlined some of the most compelling of these which directly address the open questions and priorities raised in the long range planning of astronomical communities in the UK, the USA, the EU, and Canada. Indeed, we have shown that there is an increasing call for a degree of spatial resolution from space-based platforms that, in our view, is most practically achieved through interferometry. 

Owing to the vast unexplored discovery space of the spatially resolved far-IR, the relaxed engineering requirements of these wavelengths, and the substantial history of technology maturation for this purpose, we see the development of a far-IR interferometer leveraging the double-Fourier technique as the most sensible next step for space-based interferometry and a science imperative for astronomy. Indeed, we re-affirm the conclusion of the roadmaps of previous far-IR community workshops and white papers that, following the substantial concepting and study of a large, actively cooled single-aperture facility in the form of SPICA and PRIMA, an interferometer mission that couples the sensitivity of these with sub-arcsecond angular resolution marks the next stage for far-IR astronomy. We strongly suggest that a far-IR double-Fourier interferometer mission be prioritised in the next phase of long range planning within astronomical communities.

\section*{Acknowledgements}
We thank Eleonora Alei, Steve Ertel, Joshua Lovell, Lee Mundy, David Naylor, Ettore Pedretti, Gerard van Belle, Peter Barry, and Matthew Griffin for helpful comments and discussions. 

This work has been supported by UK Research and Innovations Future Leaders Fellowship MR/W006499/1 and the Natural Science and Engineering Council --- Canada Postdoctoral Research Award. 

%%%%% References %%%%%

\bibliography{report}   % bibliography data in report.bib
\bibliographystyle{plainnat}   % makes bibtex use spiejour.bst

\end{document}